\documentclass[a4paper,10pt]{article}
\usepackage{amsfonts}
\usepackage{amsmath}
\usepackage{amssymb}
\usepackage{latexsym}
\usepackage{color}
\usepackage{colordvi}
\usepackage{epsfig}
\usepackage{array}
\usepackage{pifont}
\usepackage{cite}
\usepackage{psfrag,amsthm}
\usepackage{mathrsfs,ifpdf}

\ifpdf
\DeclareGraphicsExtensions{.pdf, .jpg}
\else
\DeclareGraphicsExtensions{.eps, .jpg}
\fi

\numberwithin{equation}{section} 
\title{\bfseries Semileptonic $\boldsymbol{B\to D^{**}}$ decays in Lattice QCD : a feasability study and first results} 

\author{M Atoui$^b$, B. Blossier$^a$,  V.~Mor\'enas$^b$\\ O.
P\`ene$^a$, K. Petrov$^c$  }

\date{}

\newcommand{\beq}{\begin{eqnarray}}
\newcommand{\eeq}{\end{eqnarray}}
\def\eq#1{Eq.~(\ref{#1})}

\usepackage{mathtools}
\usepackage{hhline}
\newcommand{\dd}{\text{d}}

\newcommand{\pD}{\ensuremath{p_{_{\!D^{**}}}}}%
\newcommand{\pB}{\ensuremath{p_{_{\!B}}}}%
\newcommand{\ED}{\ensuremath{E_{_{\!D^{**}}}}}%
\newcommand{\mD}{\ensuremath{m_{_{\!D^{**}}}}}%
\newcommand{\rD}{\ensuremath{r_{_{\!D^{**}}}}}%
\newcommand{\pL}{\ensuremath{p_{_{\!\ell}}}}%
\newcommand{\EB}{\ensuremath{E_{_{\!B}}}}%
\newcommand{\mB}{\ensuremath{m_{_{\!B}}}}%
\newcommand{\EL}{\ensuremath{E_{_{\!\ell}}}}%
\newcommand{\mL}{\ensuremath{m_{_{\!\ell}}}}%
\newcommand{\rL}{\ensuremath{r_{_{\!\ell}}}}%
\newcommand{\vpD}{\ensuremath{\vec{p}_{_{\!D^{**}}}}}%
\newcommand{\vpL}{\ensuremath{\vec{p}_{_{\!\ell}}}}%
\newcommand{\EN}{\ensuremath{E_{_{\!\nu}}}}%
\newcommand{\vpN}{\ensuremath{\vec{p}_{_{\!\nu}}}}%
\newcommand{\pN}{\ensuremath{{p}_{_{\!\nu}}}}%
\newcommand{\metr}[2]{\ensuremath{g^{#1 #2}}}%
\newcommand{\metrb}[2]{\ensuremath{g_{#1 #2}}}%

\newcommand{\poldh}[3]{\ensuremath{\varepsilon^{#1#2}_{(#3)}}}%
\newcommand{\poldhh}[3]{\ensuremath{\varepsilon^{#1#2}_{(p_{_{\!#3}},\,\lambda)}}}%
\newcommand{\poldb}[3]{\ensuremath{\varepsilon_{#1#2}^{(#3)}}}%
\newcommand{\poldbb}[3]{\ensuremath{{\varepsilon_{#1#2}^{\ast}}^{\!\!\!\!(p_{_{\!#3}},\,\lambda)}}}%

\newcommand{\rDz}{\ensuremath{r_{_{\!D^{*}_0}}}}%
\newcommand{\pDz}{\ensuremath{p_{_{\!D^{*}_0}}}}%
\newcommand{\rDd}{\ensuremath{r_{_{\!D^{*}_2}}}}%
\newcommand{\mDd}{\ensuremath{m_{_{\!D^{*}_2}}}}%
\newcommand{\pDd}{\ensuremath{p_{_{\!D^{*}_2}}}}%

\newcommand{\taub}{\ensuremath{\tau_{_{\!3/2}}}}%

\DeclarePairedDelimiter\abs{\lvert}{\rvert}   
\DeclarePairedDelimiterX\brakket[3]{\langle}{\rangle}{#1\,\delimsize\vert\,#2\,\delimsize\vert\,#3}%
\DeclarePairedDelimiterX\braket[2]{\langle}{\rangle}{#1\,\delimsize\vert\,#2}%
\DeclarePairedDelimiterX\bra[1]{\langle}{\rvert}{#1}%
\DeclarePairedDelimiterX\ket[1]{\lvert}{\rangle}{#1}%

\newcommand{\etatp}[2]{\ensuremath{\prescript{#1}{}{P}_#2}}%
\newcommand{\amplia}[3]{\ensuremath{{\mathscr T}_{#2(#3)}^#1}}%
\newcommand{\amplib}[2]{\ensuremath{{\mathscr T}_{#2}^#1}}%

\newcommand{\ampli}[3]{\ensuremath{{\mathscr T}_{#2(#3)}^#1}}%
\newcommand{\tpc}[3]{\ensuremath{A_{#1}V_{#2}D_{#3}}}%

\begin{document}

\maketitle
\begin{figure}[h]
  \begin{center}
    \includegraphics{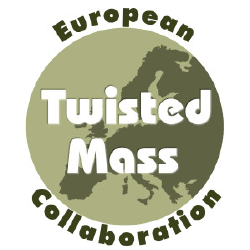}
  \end{center}
\end{figure}

\begin{center}
$^a$ Laboratoire de Physique Th\'eorique\footnote{Unit\'e Mixte de Recherche 8627 du Centre National de
la Recherche Scientifique},\\
CNRS et Universit\'e  Paris-Sud XI, B\^atiment 210, 91405 Orsay Cedex,
France\\
$^b$ Laboratoire de Physique Corpusculaire de Clermont-Ferrand\footnote{Unit\'e Mixte de Recherche 6533 CNRS/IN2P3 - Universit\'e Blaise Pascal}, Campus Universitaire des C\'ezeaux,\\
4 avenue Blaise Pascal, TSA 60026, CS 60026, 63178 Aubi\`ere Cedex, France\\
$^c$Inria Saclay, 1 rue Honor\'e d'Estienne d'Orves, B\^atiment Alan Turing, \\
Campus de l'Ecole Polytechnique, 91120 Palaiseau, France\\
\end{center}
\vspace*{20mm}

\begin{abstract}\noindent
We compute the decays ${B\to D^\ast_0}$ and 
${B\to D^\ast_2}$ with finite masses for the $b$ and $c$ quarks.
We first discuss the spectral properties of both the $B$ meson as a function of 
its momentum and of the $D^\ast_0$ and  $D^\ast_2$ at rest. We compute the 
theoretical formulae leading to the decay amplitudes from the three-point and 
two-point correlators. We then compute the amplitudes at zero recoil of 
 ${B\to D^\ast_0}$ which turns out not to be vanishing contrary to what
 happens in the heavy quark limit. This opens a possibility to get a better
 agreement with experiment. To improve the continuum limit we have
added a set of data with smaller lattice spacing.  
The ${B\to D^\ast_2}$ vanishes at zero recoil and we show a 
convincing signal but only slightly more than 1 sigma from 0. In order to reach quantitatively
significant results, we plan to fully exploit smaller lattice spacings 
as well as another lattice regularization. 

\end{abstract}

\vspace*{40mm}
\begin{flushleft}
LPC-Clermont RI 13-07\\
LPT-Orsay-13-101\\
\end{flushleft}


\newpage
\section{Introduction}

Understanding the composition of the final state in $B$ meson semileptonic decay
into charm meson is of key importance to control the theoretical error on the
CKM matrix element $V_{cb}$. The discrepancy between the inclusive determination
and the exclusive one, based on $B \to D^{(*)} l \nu$, is still of the order of
3$\sigma$~\cite{Agashe:2014kda}. A significant part of the total width $\Gamma
(B \to X_{c} l \nu)$ comes from excited states: it was recently argued that the
radial excitation $D'$ might be particularly favoured, implying a suppression of
the $B \to D^{*}$ form factors as suggested by a study performed using the
Operator Product Expansion formalism \cite{GambinoRD}. Another group of states
that contribute to the width, about one quarter of it, is orbital excitations,
in other words, positive parity charmed mesons, that we will note $D^{**}$
hereafter. They are not well understood: indeed there seems to be a persistent 
discrepancy between claims from theory and from experiment \cite{BigiQP}, while
a comparison between semileptonic decay and non leptonic decay $B \to D^{**}
\pi$, involving the same form factors (at least in the case of the so-called
Class I process), is quite confusing on the experimental side~\cite{Yaouanc:2014ypa}.
  Two types of $D^{**}$ are observed: two ``narrow resonances'' $D_{3/2}$ and a couple of
``broad resonances'' $D_{1/2}$, in the same mass region \cite{delAmoSanchezVQ}.
While experiments point towards a dominance of the broad resonances in
semileptonic decays, theory points rather towards a dominance of the narrow
resonances : not only a series of sum rules  \cite{LeYaouanc:1996bd,
Uraltsev:2000ce} derived from QCD obtains that hierarchy, but also calculations
with quark models \cite{Morenas:1997nk} - \cite{Ebert:1999ga} and lattice
computations performed in the quenched approximation \cite{Becirevic:2004ta} and
with ${\rm N_{f}}=2$ dynamical quarks \cite{BlossierVY}. However, the main
limitation of these results is that they are derived in the heavy quark
limit. $1/m_{c}$ corrections might be pretty large and, before getting any
definitive conclusion on the disagreement between theory and experiment in that
sector of flavour physics, it is mandatory to reduce the sources of systematic
errors on the theory side.

\section{Theoretical framework\label{sec:theory}}
In this paragraph, all the main formulae up to the differential decay rates will be given for the semileptonic decays of a $B$ heavy meson into the first orbitally excited $D^{**}$ mesons.\par
We will focus our study on the production of the $\ket[\big]{\etatp{3}{0}}$ (scalar $D^{*}_0$) and the $\ket[\big]{\etatp{3}{2}}$ (tensor $D^{*}_2$) states\footnote{We use the $\ket[\big]{\prescript{2S+1}{}{L}_J}$ notation of the states, where $S$ is the spin angular momentum, $L=1$ the orbital angular momentum and $J=L+S$ the total angular momentum of the $D^{**}$ state.}.\par
Finally, we will also give relations in the case where the mass of the lepton cannot be neglected.
\subsection{Form factors}
In order to derive the decay rates, we need the transition amplitudes. They can be described using 6 form factors~\cite{ISGW}.
\medskip
\par\noindent\underline{\em $\etatp{3}{2}$ state} :
\begin{equation}
\begin{aligned}\label{eq:FF1}
\brakket[\big]{\etatp{3}{2}\bigl(\pDd,\,\varepsilon(\pDd,\,\lambda)\bigr)}{V_\mu}{B(\pB)}\ &=\ i\,
\boxed{\tilde h}\,\epsilon_{\mu\nu\lambda\rho}\,
\poldhh{\ast\nu}{\alpha}{D^{*}_2}\,{\pB}_{\alpha}\,
(\pB+{\pDd})^{\lambda}\,
(\pB-{\pDd})^{\rho}\\
\brakket[\big]{\etatp{3}{2}\bigl(\pDd,\,\varepsilon(\pDd,\,\lambda)\bigr)}{A_\mu}{B(\pB)} &= \boxed{\tilde k}\,\poldbb{\mu}{\nu}{D^{*}_2}{\pB}^{\nu} +
\left({\poldbb{\alpha}{\beta}{D^{*}_2}{\pB}^{\alpha}{\pB}^{\beta}}\right)
\left[{\boxed{\tilde b_{+}}\,(\pB+\pDd)_{\mu} +
\boxed{\tilde b_{-}}\,(\pB-\pDd)_{\mu}
}\right]
\end{aligned}
\end{equation}
\par\noindent\underline{\em $\etatp{3}{0}$ state} :
\begin{equation}\label{eq:FF2}
\begin{aligned}
\brakket[\big]{\etatp{3}{0}(\pDz)}{V_\mu}{B(\pB)} &=0\qquad\quad\text{(parity invariance)}\\
\brakket[\big]{\etatp{3}{0}(\pDz)}{A_\mu}{B(\pB)}&=
\boxed{\tilde u_+}\,(\pB+\pDz)_{\mu} + 
\boxed{\tilde u_-}\,(\pB-\pDz)_{\mu}
\end{aligned}
\end{equation}
where $V_\mu$ denotes the vector current $\bar c\gamma_\mu b$ and $A_\mu$ the axial current $\bar c\gamma_\mu\gamma_5 b$.\par
$\varepsilon(\pDd,\,\lambda)$ is the polarisation tensor of the \etatp{3}{2} state ($\lambda$ being the projection of the $J=2$ total angular momentum along some quantification axis).\par
Moreover, the chosen normalisation of the mesonic states is
\[
\braket[\big]{M\bigl(p^{\,\prime}\bigr)}{M\bigl(p\bigr)}\ =\ (2\pi)^3\,2E\,\delta^3(\vec p^{\,\prime} 
- \vec p)\,.
\]
Finally, because of parity and time-reversal invariance of the strong interactions, those form factors are real numbers.
\subsection{Differential decay rates}
The goal is to compute the differential decay width $\dd\Gamma(\bar B\to D^{**}\,\ell\,\bar\nu)$ whose general expression is
\begin{gather*}
\dd\Gamma(\bar B\to D^{**}\,\ell\,\bar\nu) = 
\dfrac{1}{2\,\EB}\,\abs{\bar{\mathscr{M}}}^2\,\dd\Phi\,,\\[2mm]
\text{with}\qquad\qquad
\left\{
\begin{aligned}
&\dd\Phi= 
\dfrac{d^3\vpD}{(2\pi)^3\,2\ED}\,\dfrac{d^3\vpL}{(2\pi)^3\,2\EL}\,\dfrac{d^3\vpN}{(2\pi)^3\,2\EN}\,(2\pi)^4\,\delta^{(4)}(\pB-\pD-\pL-\pN)\label{eq:gamma}\,,\\[2mm]
&\abs{\bar{\mathscr{M}}}^2 = \sum\limits_{\mu,\,\nu}\,W_{\mu\nu}\ell^{\mu\nu}\,.
\end{aligned}
\right.\notag
\end{gather*}
In the last equality, $W_{\mu\nu}$ denotes the hadronic tensor
\[
W_{\mu\nu}(\pB,\,\pD) = \dfrac{G_F^2\,\abs{V_{cb}}^2}{2}\,\sum_{\mathclap{{\text{\tiny final spins}}}}\ 
\brakket[\big]{D^{\ast\ast}(\pD)}{V_{\mu} - A_{\mu}}{\bar B(\pB)}\ 
\brakket[\big]{\bar B(\pB)}{V_{\nu} - A_{\nu}}{D^{\ast\ast}(\pD)}\,,
\]
where the transition amplitudes have been given in the preceding paragraph (let us note that there are no summation nor average over the initial spins since the $\bar B$ meson has a spin equal to zero) and $\ell^{\mu\nu}$ represents the leptonic tensor
\begin{equation*}
\ell^{\mu\nu}(\pL,\,\pN) = \sum_s\left[{
\bar u_\ell(\pL,s)\,\gamma^\mu\!\left({1 - \gamma^5}\right)\!v_\nu(\pN)
}\right]\!\!\cdot\!\!
\left[{
\bar u_\ell(\pL,s)\,\gamma^\nu\!\left({1 - \gamma^5}\right)\!v_\nu(\pN)
}\right]^\ast 
\end{equation*}
In that last formula, $u_\ell(\pL,s)$ is the lepton $\ell$ spinor ($s$ denotes the usual projection of its spin), while $v_\nu(\pN)$ represents the antineutrino $\bar\nu$ spinor.\par\noindent
All that remains is to compute the leptonic tensor, then the hadronic tensor and the measure $\dd\Phi$ of the phase space in order to obtain the expressions of the differential decay widths.
\subsubsection{Leptonic tensor $\boldsymbol{\ell^{\mu\nu}}$}
The calculation is classical and straightforward, leading to
\begin{equation*}
\ell^{\mu\nu} = 8\left[{
\pL^\mu\,\pN^\nu\ +\ \pL^\nu\,\pN^\mu\ - (\pL\cdot\pN)\,\metr{\mu}{\nu}\ -\ i\,\epsilon^{\mu\nu\rho\sigma}(\pL)_\rho(\pN)_\sigma
}\right]\,.
\end{equation*}
We can notice that the mass of the lepton has vanished, which renders the expression valid in the situations where $\mL = 0$ as well as $\mL\neq 0$.
\subsubsection{Hadronic tensor $\boldsymbol{W_{\mu\nu}}$}
By looking at the expressions of the transition amplitudes given above, the general structure of the hadronic tensor can be inferred and put into the form~\cite{ISGW}:
\begin{equation}\label{eq:hadro}
\begin{split}
W_{\mu\nu} &= \dfrac{G_F^2\,\abs{V_{cb}}^2}{2}\,\biggl[{
\boxed{\alpha}\,\metrb{\mu}{\nu}\ +\ \boxed{\beta_{_{++}}}(\pB+\pD)_\mu(\pB+\pD)_\nu\ +\ \boxed{\beta_{_{+-}}}(\pB+\pD)_\mu(\pB-\pD)_\nu}\biggr.\\
&\biggl.{\ +\ \boxed{\beta_{_{-+}}}(\pB-\pD)_\mu(\pB+\pD)_\nu\ +\ \beta_{_{--}}(\pB-\pD)_\mu(\pB-\pD)_\nu\ +\ i\,\boxed{\gamma}\,\epsilon_{\mu\nu\rho\sigma}(\pB+\pD)^\rho(\pB-\pD)^\sigma
}\biggr]\,.
\end{split}
\end{equation}
The coefficients ${\alpha}$, ${\beta_{_{++}}}$, ${\beta_{_{+-}}}$, ${\beta_{_{-+}}}$, 
${\beta_{_{--}}}$ and ${\gamma}$ are given in the Appendix for the \etatp{3}{0} and the \etatp{3}{2} 
states. \subsubsection{Kinematics and notations}
For reasons of simplification, we now choose to compute the decay rates in the rest frame of the $\bar B$ meson.\par 
We then define two dimensionless parameters $x$ and $y$ according to
\begin{center}
\begin{minipage}{40mm}
\begin{equation*}
\boxed{x\,\mB = 2\,\EL}
\end{equation*}
\end{minipage}
\begin{minipage}{100mm}
\begin{equation*}
\text{as well as}\qquad\qquad\qquad\boxed{y\,\mB^2 = (\pB-\pD)^2 = (\pL+\pN)^2}
\end{equation*}
\end{minipage}
\end{center}
where $\EL$ is the energy of the lepton in the $\bar B$ rest frame.\par
We introduce also the mass ratio $r_{_{\!X}}$
\[
\boxed{
m_{_{\!X}}\ =\ r_{_{\!X}}\,{\mB}
}\qquad\text{where $X$ is either a $D^{**}$ meson or the lepton $\ell$}
\]
Many kinematical terms can be expressed with these three parameters, such as
\[
\ED = \dfrac{\mB}{2}\,(1 - y + \rD^2)\quad;\quad \pL\cdot\pN = \dfrac12\,\mB^2\,(y - \rL^2)\quad;\quad
\pL\cdot\pD = \dfrac12\,\mB^2\,(x -y - \rL^2)\quad;\quad \pB\cdot\pL = \dfrac12\,\mB^2\,x \quad.
\]
\subsubsection{Measure $\boldsymbol{\dd\Phi}$ of the phase space}
The goal is to get the differential widths $\dd\Gamma$ with respect to the lepton energy $\EL$ and the momentum transfer $(\pB-\pD)^2$, in other words with respect to the variables $x$ and $y$ : $\ {\dd^2\Gamma}/{\dd x\,\dd y}$\par
So we must integrate over the antineutrino momentum $\vpN$, then over all possible orientations of $\vpL$ so that only the dependance on $\EL$ (i.e. on $x$) remains, and finally over all possible directions of the 3-vector $\vpD$ since we want to keep the dependance on $\ED$ (i.e. on $y$). We finally get :
\[
\dd\Phi = -\,\dfrac{\mB^2}{128\pi^3}\,\dd x\,\dd y\,\theta(1 - x + y - \rD^2)
\]
where $\theta(z)$ is the usual Heaviside function.
\subsubsection{Constraints on $\boldsymbol{x}$ and $\boldsymbol{y}$}
The parameters $x$ and $y$, that is the lepton energy ($\EL$) and the $D^{\ast\ast}$ meson energy 
($\ED$), cannot be arbitrary. They are constrained by two conditions : one which is obvious in the 
expression of $\dd\Phi$ above (the Heaviside function) and another one which appeared during the 
integration over the direction of $\vpL$. In other terms, we have access to the variation domains of 
both parameters $x$ and $y$ whether we consider $x=x(y)$ or $y=y(x)$: they are given in the Appendix.
\subsubsection{Differential decay widths in the $\boldsymbol{\bar B}$ rest frame}
Using the definition of $\dd\Gamma$ as well as all the preceding results, the construction of the differential decay widths proceeds in the following way :
\[
\boxed{
\dfrac{\dd\Gamma}{\dd x\,\dd y}(\bar B\to D^{**}\,\ell\,\bar\nu) = -\,\dfrac{\mB}{256\pi^3}\,\abs{\bar{\mathscr{M}}}^2
}
\]
where $\abs{\bar{\mathscr{M}}}^2 = W_{\mu\nu}\ell^{\mu\nu}$ becomes in this particular frame
\[
\begin{aligned}
\abs{\bar{\mathscr{M}}}^2 =
2\,{G_F^2\,\abs{V_{cb}}^2}\,\mB^2\,\Biggl\{{}\Biggr. &-2\,\boxed{\alpha}\,(y - \rL^2)\\
&- \boxed{\beta_{_{++}}}\,\mB^2\,\Bigl[{4\,\bigl[{x\,\rD^2 + (1-x)(y-x)}\bigr]+\rL^2\bigl[{3\,y -4(x+\rD^2)+\rL^2}\bigr]}\Bigr]\\[3mm]
&+ \left({\,\boxed{\beta_{_{+-}}} + \boxed{\beta_{_{-+}}}\,}\right)\,\mB^2\,\rL^2\Bigl[{2(1-x-\rD^2) + y + \rL^2}\Bigr]\\[3mm]
&+ \boxed{\beta_{_{--}}}\,\mB^2\,\rL^2\,(y - \rL^2)\\
&- 2\,\boxed{\gamma}\,\mB^2\Bigl[
{y\,(1+y-2\,x-\rD^2) + \rL^2(1+y-\rD^2)}
\Bigr]\Biggl.{}\Biggr\}
\end{aligned}
\]
We can notice that, for a zero mass lepton, only the coefficients ${\alpha}$, ${\beta_{_{++}}}$ and ${\gamma}$ survive.\par
The expressions for each $D^{**}$ are also written in the Appendix.
However, their use requires the knowledge of the momentum dependance of the form factors. In the following, we will focus on a method to obtain such a dependance.
\subsection{Extracting the form factors from the transition amplitudes }
On the lattice, we compute the transition amplitudes for different momenta of the mesons. But we need the momentum dependance of the form factors in order to calculate the decay rates of the semileptonic decays of the $B$ to a $D^{**}$. So we must devise a way to extract the form factors from the lattice transition amplitudes.
\subsubsection{Kinematics}
We will work in the rest frame of the $D^{**}$ meson\footnote{This will greatly simplify the calculations on the lattice.} so the $B$ meson will carry the momentum. Moreover, we will consider the $B$'s whose spatial momentum is symmetrical.
\[
\pD = (\mD,\,\vec 0)\quad\quad\text{and}\quad\quad \pB^\mu = (\EB,\,p,\,p,\,p)\quad.
\]
We will also choose the Minkowski metrics: $\ g_{\mu\nu} = \text{Diag}(+,\,-,\,-,\,-)$\par\noindent
The other piece we need is the expression of the polarisation tensor for the \etatp{3}{2} state in the $D^{**}$ rest frame, that is
$\varepsilon(\vec 0,\,\lambda)$. We can construct it from the combination of two spin-1 states
\[
{\varepsilon^{\mu\nu}(\vec 0,\,\lambda) = \sum\limits_{s,\,s^\prime}\braket{1\ 1\ s\ s^\prime}{2\ 
\lambda}\,\varepsilon^{\mu}(\vec 0,\,s)\,\varepsilon^{\nu}(\vec 0,\,s^\prime)}\,,
\]
where the Clebsch-Gordan coefficients for $1 + 1 \to 2$ appear, as well as the polarisation vector 
$\varepsilon^{\mu}(\vec 0,\,s)$ of a spin-1 state. The final expressions are gathered in the 
Appendix. \subsubsection{$\boldsymbol{\etatp{3}{0}}$ form factors}
Using the notation
\[
\amplib{A}{\mu}\overset{\text{def.}}{=} \brakket[\big]{\etatp{3}{0}}{A_\mu}{B(\pB)},
\]
we explicitely get from~Eqs.~\eqref{eq:FF2}:
\begin{equation*}
\left\{
\begin{aligned}
\amplib{A}{0} &= \boxed{\tilde u_+}\,(\EB+\mD) + \boxed{\tilde 
u_-}\,(\EB-\mD)&\qquad\qquad\text{(temporal direction),}\\
\amplib{A}{i} &= \boxed{\tilde u_+}\,p + \boxed{\tilde u_-}\,p&\qquad\qquad\text{(spatial 
direction).}
\end{aligned}
\right.
\end{equation*}
So it is straightforward to express $\tilde u_+$ and $\tilde u_-$ with the $\amplib{A}{\mu}$'s. The 
results are presented in the Appendix.
\subsubsection{$\boldsymbol{\etatp{3}{2}}$ form factors}\label{3P2formfactors}
In the following, we will adopt the notation:
\[
\amplia{A}{\mu}{\lambda}\overset{\text{def.}}{=} \brakket[\big]{\etatp{3}{2}(\lambda)}{A_\mu}{B(\pB)}\qquad\text{as well as}\qquad
\amplia{V}{\mu}{\lambda}\overset{\text{def.}}{=} 
\brakket[\big]{\etatp{3}{2}(\lambda)}{V_\mu}{B(\pB)}.
\]
In order to extract one particular form factor, we can choose in~Eqs.~\eqref{eq:FF1} either some spatial direction 
where each coefficient of the other form factors vanishes, or we can construct a linear combination of 
the $\amplia{A}{i}{\lambda}$ and/or the $\amplia{V}{i}{\lambda}$.\par\noindent
This procedure can be carried out by using the expressions for the polarisation tensor and the four-momenta at our disposal and calculating the contribution of the corresponding terms appearing in the matrix elements~\eqref{eq:FF1} which define the form factors (those contributions are gathered in Table~\ref{tab:un}).
\par\noindent
\begin{table}[htb]
\begin{center}
\begin{tabular}{||c|c||}
\hhline{|t:==:t|}
\rule{0pt}{14pt}$\boldsymbol{\poldb{\mu}{\nu}{\lambda}}$ & $\boldsymbol{{\poldb{\mu}{\nu}{\lambda}\pB^{\nu}}}$\\[5pt]
\hhline{|:=:=:|}
\rule{0pt}{15pt}$\poldb{\mu}{\nu}{+2}$ & $\dfrac{p}{2}\,\Bigl({0,\,1 + i,\,i - 1,\,0}\Bigr)$\\[5pt]
\hhline{||-|-||}
\rule{0pt}{15pt}$\poldb{\mu}{\nu}{+1}$ & $-\,\dfrac{p}{2}\,\Bigl({0,\,1,\,i,\,1 + i}\Bigr)$\\[5pt]
\hhline{||-|-||}
\rule{0pt}{15pt}$\poldb{\mu}{\nu}{0}$ & $-\,\dfrac{p}{\sqrt{6}}\,\Bigl({0,\,1,\,1,\,-\,2}\Bigr)$\\[6pt]
\hhline{||-|-||}
\rule{0pt}{15pt}$\poldb{\mu}{\nu}{-1}$ & $\dfrac{p}{2}\,\Bigl({0,\,1,\,-\,i,\,1 - i}\Bigr)$\\[5pt]
\hhline{||-|-||}
\rule{0pt}{15pt}$\poldb{\mu}{\nu}{-2}$ & $\dfrac{p}{2}\,\Bigl({0,\,1 - i,\,-\,1 -\,i,\,0}\Bigr)$\\[5pt]
\hhline{|:=:=:|}
\rule{0pt}{14pt}$\poldb{\mu}{\nu}{+2} + \poldb{\mu}{\nu}{-2}$ &  $p\,\Bigl({0,\,1 ,\,-\,1,\,0}\Bigr)$\\[5pt]
\hhline{||-|-||}
\rule{0pt}{14pt}$\poldb{\mu}{\nu}{+2} - \poldb{\mu}{\nu}{-2}$ &  $p\,\Bigl({0,\,i ,\, i,\,0}\Bigr)$\\[5pt]
\hhline{||=|=||}
\rule{0pt}{14pt}$\poldb{\mu}{\nu}{+1} + \poldb{\mu}{\nu}{-1}$ &  $p\,\Bigl({0,\,0,\,-\,i,\,-\,i}\Bigr)$\\[5pt]
\hhline{||-|-||}
\rule{0pt}{14pt}$\poldb{\mu}{\nu}{+1} - \poldb{\mu}{\nu}{-1}$ &  $p\,\Bigl({0,\,-\,1 ,\,0,\,-\,1}\Bigr)$\\[5pt]
\hhline{|b:==:b|}
\end{tabular}$\qquad$
\begin{tabular}{||c|c||}
\hhline{|t:==:t|}
\rule{0pt}{14pt}$\boldsymbol{\poldh{\mu}{\nu}{\lambda}}$ & $\boldsymbol{{\poldh{\mu}{\nu}{\lambda}{\pB}_{\nu}}}$\\[5pt]
\hhline{|:=:=:|}
\rule{0pt}{15pt}$\poldh{\mu}{\nu}{+2}$ & $-\,\dfrac{p}{2}\,\Bigl({0,\,1 + i,\,-\,1 + i,\,0}\Bigr)$\\[5pt]
\hhline{||-|-||}
\rule{0pt}{15pt}$\poldh{\mu}{\nu}{+1}$ & $\dfrac{p}{2}\,\Bigl({0,\,1,\,i,\,1 + i}\Bigr)$\\[5pt]
\hhline{||-|-||}
\rule{0pt}{15pt}$\poldh{\mu}{\nu}{0}$ & $\dfrac{p}{\sqrt{6}}\,\Bigl({0,\,1,\,1,\,-\,2}\Bigr)$\\[6pt]
\hhline{||-|-||}
\rule{0pt}{15pt}$\poldh{\mu}{\nu}{-1}$ & $\dfrac{p}{2}\,\Bigl({0,\,-\,1,\,i,\,-\,1 + i}\Bigr)$\\[5pt]
\hhline{||-|-||}
\rule{0pt}{15pt}$\poldh{\mu}{\nu}{-2}$ & $\dfrac{p}{2}\,\Bigl({0,\,-\,1 + i,\,1 + i,\,0}\Bigr)$\\[5pt]
\hhline{|:=:=:|}
\rule{0pt}{14pt}$\poldh{\mu}{\nu}{+2} + \poldh{\mu}{\nu}{-2}$ &  $p\,\Bigl({0,\,-\,1,\,1,\,0}\Bigr)$\\[5pt]
\hhline{||-|-||}
\rule{0pt}{14pt}$\poldh{\mu}{\nu}{+2} - \poldh{\mu}{\nu}{-2}$ &  $p\,\Bigl({0,\,-\,i,\,-\,i,\,0}\Bigr)$\\[5pt]
\hhline{||=|=||}
\rule{0pt}{14pt}$\poldh{\mu}{\nu}{+1} + \poldh{\mu}{\nu}{-1}$ &  $p\,\Bigl({0,\,0,\,i,\,i}\Bigr)$\\[5pt]
\hhline{||-|-||}
\rule{0pt}{14pt}$\poldh{\mu}{\nu}{+1} - \poldh{\mu}{\nu}{-1}$ &  $p\,\Bigl({0,\,1 ,\,0,\,1}\Bigr)$\\[5pt]
\hhline{|b:==:b|}
\end{tabular}
\end{center}\par\noindent\bigskip
\begin{center}
\begin{tabular}{||c|c||}
\hhline{|t:==:t|}
\rule{0pt}{14pt}$\boldsymbol{\poldh{\mu}{\nu}{\lambda}}$ & $\boldsymbol{{\poldh{\mu}{\nu}{\lambda}{\pB}_{\mu}{\pB}_{\nu}}}$\\[5pt]
\hhline{|:=:=:|}
$\poldh{\mu}{\nu}{+2}$ & $i\,p^2$\\[2pt]
\hhline{||-|-||}
$\poldh{\mu}{\nu}{+1}$ & $-\,(1 + i)\,p^2$\\[2pt]
\hhline{||-|-||}
$\poldh{\mu}{\nu}{0}$ & $0$\\[2pt]
\hhline{||-|-||}
$\poldh{\mu}{\nu}{-1}$ & $(1 - i)\,p^2$\\[2pt]
\hhline{||-|-||}
$\poldh{\mu}{\nu}{-2}$ & $-\,i\,p^2$\\[2pt]
\hhline{|:=:=:|}
$\poldh{\mu}{\nu}{+2} + \poldh{\mu}{\nu}{-2}$ &  $0$\\[2pt]
\hhline{||-|-||}
$\poldh{\mu}{\nu}{+2} - \poldh{\mu}{\nu}{-2}$ &  $2\,i\,p^2$\\[2pt]
\hhline{|:=:=:|}
$\poldh{\mu}{\nu}{+1} + \poldh{\mu}{\nu}{-1}$ &  $-\,2\,i\,p^2$\\[2pt]
\hhline{||-|-||}
$\poldh{\mu}{\nu}{+1} - \poldh{\mu}{\nu}{-1}$ &  $-\,2\,p^2$\\[2pt]
\hhline{|b:==:b|}
\end{tabular}
\end{center}
\caption{\label{tab:un}\em Contributions of the polarisation tensor in the $B\,\to\,\etatp{3}{2}$ transition amplitude}
\end{table}
A few possibilities are collected in the Appendix.
\subsection{Summary}
We have constructed all the theoretical formulae which allow us to calculate the decay widths of the semileptonic $B \to D^{**}$ channels. The strategy to use them is the following :
\begin{enumerate}
\item compute, on the lattice, the transition amplitudes for the $B \to D^{**}$ processes.
\item extract the form factors from them.
\item use the formulae in the Appendix to obtain the decay widths.
\end{enumerate}
Since we expect the lattice \etatp{3}{2} computation to be somewhat tricky, we are first going to estimate the contribution of the $\tilde k$, $\tilde b_+$, $\tilde b_-$ and $\tilde h$ form factors to the $\bar B \to D^*_2\,\ell\,\bar\nu$ decay width.
\subsection{Estimation of the contribution of the form factors to the $\boldsymbol{\etatp{3}{2}}$ decay width}\label{sec:estimation}
There are four form factors needed to describe the transition amplitudes from a $B$ to a \etatp{3}{2} state which increases the difficulty in the lattice computations. So it could be useful to have an idea of each of their contribution to the decay widths.\par\noindent
In order to get a quantitative hint, we will relate these form factors to their infinite mass limit \taub$\,$ and use this \taub$\,$ to produce a numerical estimation.
\subsubsection{Infinite mass limit}\noindent
In the limit where the heavy quark of the meson has an infinite mass, new symmetries (and thus additionnal conserved quantities) appear. These new symmetries provide additional relations between the transition amplitudes so that the form factors become dependant. It can be proven~\cite{ISGW2} that this reduction of the form factors leads to the following relations for the \etatp{3}{2} state :
\[
\left\{
\begin{aligned}
&\quad{{\tilde h} = \dfrac{\sqrt{3}}{2}\dfrac{1}{\mB^2\,\sqrt{\rDd}}\,\taub}&\qquad\qquad\qquad&{{\tilde k} = \sqrt{3}\,\sqrt{\smash[b]{\rDd}}\,(1+w)\,\taub}\\[2mm]
&\quad{{\tilde b}_+ = -\,\dfrac{\sqrt{3}}{2}\dfrac{1}{\mB^2\,\sqrt{\rDd}}\,\taub}&\qquad\qquad\qquad&{{\tilde b}_- = \dfrac{\sqrt{3}}{2}\dfrac{1}{\mB^2\,\sqrt{\rDd}}\,\taub}
\end{aligned}
\right.
\]
where the parameter $w$ is defined by :
\[
\mB\,\mDd\,w = \pB\!\cdot\pDd\qquad\Longrightarrow\qquad\boxed{y = 1 + \rD^2 - 2\,\rD\,w}
\]
and \taub$\,$ is one of the so-called Isgur-Wise functions.
\subsubsection{Fit of $\boldsymbol{\taub}$}\label{sec:fit}\noindent
Using a covariant construction of the transition amplitudes in the infinite mass limit (quark models \`a la Bakamjian-Thomas), it has been shown~\cite{Bakamjian:1953kh,Morenas:1997nk} that the Isgur-Wise function \taub$\,$ can be well fitted by :
\[
{\taub(w) = \taub(1)\,\left(\dfrac{2}{1+w}\right)^{2\,\sigma^2_{_{\!3/2}}}}\qquad\Longrightarrow\qquad
\taub(y) = \taub(1)\,\left[\dfrac{4\,\rDd}{(1+\rDd)^2-y}\right]^{2\,\sigma^2_{_{\!3/2}}}
\]
where the accessible phase space domain is given by :
\[
1\le w\le \dfrac{\mB^2+\mD^2}{2\,\mB\,\mD}\qquad\Longrightarrow\qquad{(1-\rDd)^2\ge y\ge 0}
\]
We will also take (GI model~\cite{ISGW} in~\cite{Morenas:1997nk}) : $\qquad\qquad{\taub(1)\simeq 
0.54}\quad\text{as well as}\quad{\sigma^2_{_{\!3/2}} \simeq  1.50}\,.$
\subsubsection{Quantitative prediction of each contribution to the total width}\noindent
We are now in position to estimate the contribution of each form factor to the total width of the 
$\bar B\to D^{*}_2\,\ell\,\bar\nu$ decay channel. Let us take the case of a zero mass lepton to 
simplify the calculations. Starting from the expression of $\dfrac{\dd^2\Gamma}{\dd x\,\dd y}$ and 
with the notations given in the Appendix, we can perform both the integrations over $x$ and $y$ and 
we get : \begin{center}
\begin{tabular}{||c||c|c|c|c|c||}
\hhline{|t:=:t:=====:t|}
\rule{0pt}{12pt}
$\boldsymbol{C_i}$&$C_1\times{\tilde k}^2$&$C_2\times{\tilde h}^2$&$C_3\times{\tilde b}_+^2$&$C_5\times2\,{\tilde k}\,{\tilde b}_+$&$C_8$\\[2pt] \hhline{|:=::=:=:=:=:=:|}
\rule{0pt}{12pt}
$\boldsymbol{\iint C_i\times\text{\bfseries FF}^2}$&-61.3&-0.86&-4.43&29.0&0 \\[2pt]
\hhline{|b:=:b:=====:b|}
\end{tabular}
\end{center}
We can notice that the biggest contributions come from the terms where the ${\tilde k}$ form factor appears; that is why we will focus on its determination in the actual lattice computation.

\section{Simulation set up\label{sec:simulation}}

In our analysis we use gauge ensembles produced by European Twisted Mass 
Collaboration \cite{Boucaud:2007uk} - \cite{Boucaud:2008xu} with $N_f=2$ twisted-mass
fermions tuned at maximal twist. Parameters of the simulations are collected
in Table \ref{tabsimul}.
\begin{table}[btph]
\begin{center}
\begin{tabular}{||c|c|c|c|c|c|c|c||}
\hhline{|t:========:t|}
$\boldsymbol{\beta}$&$\boldsymbol{L^3\times T}$&$\boldsymbol{a}$\bfseries [fm]&\bfseries\# cnfgs&$\boldsymbol{\mu_{\rm sea}=\mu_l}$&$\boldsymbol{\mu_c}$&
$\boldsymbol{\mu_b}$&$\boldsymbol{\theta\ [\pi/L]}$\\
\hhline{|:=:=:=:=:=:=:=:=:|}
3.9&$24^3 \times 48$&$0.085(3)$&240&0.0085&0.215&0.3498&0.0, 0.99, 1.41\\
&&&&&&&2.02,\,2.50,\,2.92\\
&&&&&&&3.66\\
\hhline{||~|~|~|~|~|~|-|-||}
&&&&&&0.4839&0.0,1.21,1.72\\
&&&&&&&2.46,\,3.05,\,3.56\\
&&&&&&&4.46\\
\hhline{||~|~|~|~|~|~|-|-||}
&&&&&&0.6694&0.0,1.48,2.11\\
&&&&&&&3.01,\,3.73,\,4.36\\
&&&&&&&5.46\\
\hhline{|:=:=:=:=:=:=:=:=:|}
4.05&$32^3\times 64$&0.069(2)&160&0.006&0.1849&0.3008&0.0,1.09,1.56\\
&&&&&&&2.23,\,2.76,\,3.23\\
&&&&&&&4.04\\
\hhline{||~|~|~|~|~|~|-|-||}
&&&&&&0.4162&0.0,1.35,1.92\\
&&&&&&&2.74,\,3.40,\,3.97\\
&&&&&&&4.97\\
\hhline{||~|~|~|~|~|~|-|-||}
&&&&&&0.5757&0.0,1.67,2.37\\
&&&&&&&3.39,\,4.21,\,4.91\\
&&&&&&&6.15\\
\hhline{|:=:=:=:=:=:=:=:=:|}
4.2&$32^3\times 64$&0.054(2)&300&0.0065&0.1566&0.2548&0.0\\
&&&&&&0.3525&0.0\\
&&&&&&0.4876&0.0\\
\hhline{|b:========:b|}
\end{tabular}
\end{center}
\caption{\label{tabsimul}\em Parameters of the simulations used in this work; masses and
momenta are expressed in lattice units. Pion masses are $m_\pi=420$ MeV 
for the ensemble ($\beta=3.9$, $a\mu_{\rm sea}=0.0085$) and $m_\pi=450$ 
MeV for the ensemble ($\beta=4.05$, $a\mu_{\rm sea}=0.0060$) 
\cite{Blossier:2010cr}.
The lattice spacing $a_{\beta=3.9}$ is fixed by imposing the matching of $f_\pi$ 
obtained on the lattice to the experimental value \cite{Blossier:2010cr} 
and $a_{\beta=4.05}$ is rescaled
using the parameter $\Lambda^{\rm N_f=2}_{\overline{\rm MS}}$ \cite{Blossier:2010ky}.
We have added a premiminary use of data for ($\beta=4.2$, $a\mu_{\rm sea}=0.0065$)
and $m_\pi=495$ MeV. This will be used only for the decay into a scalar charmed
meson. The reason is that, with only $\beta=3.9$ and $\beta=4.05$,
the extrapolation of the decay amplitude $B \to D^\star_0 l \nu$ to the physical
situation produces a result grieved by more than 100 \% error. The data for
$\beta=4.2$, being closer to the coninuum, allow a significant result as will be
seen.}
\end{table}
 The gauge action is tree-level Symanzik improved 
\cite{Weisz:1982zw} and reads
\begin{equation*}
S_\mathrm{G}[U] = \frac{\beta}{6} \bigg(b_0 \sum_{x,\mu\neq\nu} 
\textrm{Tr}\Big(1 - P^{1 \times 1}(x;\mu,\nu)\Big) + b_1 \sum_{x,\mu\neq\nu} \textrm{Tr}
\Big(1 - P^{1 \times 2}(x;\mu,\nu)\Big)\bigg) ,
\end{equation*}
where $b_0 = 1 - 8 b_1$ and $b_1 = -1/12$. The fermionic action with two degenerate 
flavors is Wilson-like
with a twisted mass term and reads 
\cite{Frezzotti:2000nk} - \cite{Shindler:2007vp}:
\begin{equation*}
S_\mathrm{F}[\chi_q,\bar{\chi}_q,U] = a^4 \sum_x \bar{\chi}_q(x) 
\Big(D_{\rm W} + i\mu_\mathrm{q}\gamma_5\tau_3\Big) \chi(x)_q , 
\end{equation*}
where $D_{\rm W}$ is the massless Wilson-Dirac operator.
In the valence sector we add two doublets of charm quarks and ``bottom'' quarks. Moreover, as we 
are interested in computing form factors at different momenta 
we implement $\theta$-boundary conditions~\cite{Jansen:1995ck}, using 
$\vec{\theta}\equiv (\theta,\theta,\theta)$, 
for the $b$ doublet: $$\chi_b(x + L\hat{e}_i)=e^{i \theta L}
\chi_b(x)$$
This is equivalent to define an auxiliary field
$$\chi^{\vec{\theta}}_b(x) = e^{-i {\vec{\theta}}\cdot \vec{x}} \chi_b(x)$$ and a Dirac operator
\[
D^{\vec{\theta}}(\chi_b, \bar{\chi}_b, U) \equiv D(\chi^{\vec{\theta}}_b, 
\bar{\chi}^{\vec{\theta}}_b, 
U^{\vec{\theta}})\qquad\text{with}\qquad U^{\vec{\theta}}_i(x) =e^{i a \theta} U_i(x).
\]
The whole fermionic action reads finally :
\[
S^{\rm val}\ =\ S_\mathrm{F}[\chi_q,\,\bar{\chi}_q,\,U]\ +\ 
S_\mathrm{F}[\chi_c,\,\bar{\chi}_c,\,U]\ +\  
S_\mathrm{F}\bigl[\chi^{\vec{\theta}}_b,\,\bar{\chi}^{\vec{\theta}}_b,\,U^{\vec{\theta}}\bigr].
\]
We use all to all propagators with stochastic sources $\eta[i]$ diluted in time
\cite{Foley:2005ac} and improve the variance to signal ratio with the one-end trick
\cite{Foster:1999wu,McNeile:2006bz}. When it is generalised to $\theta$-boundary conditions, it
consists in solving the Dirac equations
\begin{equation*}
\sum_{y} D\bigl[f,\,r,\,\vec{\theta}\,\bigr]^{ab}_{\alpha \beta}(x,y)\ 
\phi\bigl[i,\,f,\,r,\,\vec{\theta},\,\tilde{\alpha},\,\tilde{t}\,\bigr]^b_{\beta}(y)\ =\ 
\eta[i]^a_\alpha(x)\ \delta_{\alpha\,\tilde{\alpha}}\ \delta_{t_x\,\tilde{t}}\,,
\end{equation*}
where $\tau^3 \chi = r \chi$, $f$ represents the fermion flavour, and
\begin{equation*}
    \sum_{y} D\bigl[f_2,\,r_2,\,\vec{\theta}_2\bigr]^{ab}_{\alpha\beta}(x, y) \  
    \Phi\bigl[i,\,f_2,\,r_2,\,f_1,\,r_1,\,\Gamma_2,\,\vec{\theta}_2,\,\vec{\theta}_1,\,\tilde{\alpha},\,
    \tilde{t},\,\tilde{t}+t_S\bigr]_{\beta}^b(y)
 \  =\  \Gamma_{2\,\alpha\,\beta} \ 
    \phi\bigl[i,\,f_1,\,r_1,\,\vec{\theta}_1,\,\tilde{\alpha},\,\tilde{t}\,\bigr]_{\beta}^a(x) 
    \  \delta_{t_S,\, t_x - \tilde{t}}\,.
\end{equation*}
The stochastic source $$\xi\bigl[i,\,\tilde{\alpha},\,\tilde{t}\,\bigr]^a_{\alpha}(x)\ \equiv\ 
\eta[i]^a_\alpha(x)ó\delta_{\alpha\,\tilde{\alpha}}\ \delta_{t_x\,\tilde{t}}$$ is diluted in spinor 
and 
is non zero in a single time-slice $\tilde{t}$. It is normalized by
\[
\lim_{N \to \infty} \dfrac{1}{N} \sum_{i=1}^N 
\xi\bigl[i,\,\tilde{\alpha},\,\tilde{t}\,\bigr]^a_{\alpha}(x)\ 
\xi^*\bigl[i,\,\tilde{\alpha},\,\tilde{t}\,\bigr]^{b}_{\beta}(y)\ =\ \delta_{ab}\,
\delta_{\alpha\beta}\,\delta_{xy}\,\delta_{\alpha
\tilde{\alpha}}\delta_{t_x\tilde{t}}\,.
\]
In order to improve the overlap of the interpolating fields for the ground states or 
to create operator of higher spin (for instance the tensor meson $D^*_2$), one
has to use interpolating fields generically written as 
$\bar{\chi}_1 S \times \Gamma \chi_2$, where $S$ is a path of links and $\Gamma$
is any Dirac matrix. We use interpolating fields of the so-called Gaussian 
smeared-form~\cite{Gusken:1989ad}
\beq\nonumber
S = \left(\frac{1 + \kappa_G a^2 \Delta}{1+6\kappa_G}\right)^{R},
\eeq
where $\kappa_G=0.15$ is a hopping parameter, $R=30$ is the number of 
applications of the operator $(1 + \kappa_G a^2 \Delta)/(1+6\kappa_G)$, and $\Delta$ the 
gauge-covariant 3-D Laplacian constructed from three-times APE-blocked 
links~\cite{Albanese:1987ds}. If necessary, we also incorporate in $S$ a covariant derivative: 
$$\nabla_i \equiv \frac{1}{2a} \left[U_i(x) - U^\dag_i(x - \hat{\imath})\right].$$
It is the case to create a tensor meson.\par
The Dirac equations, which we then have to solve, read:
\[
\left\{
\begin{aligned}
&\sum_{y} D\bigl[f,\,r,\,\vec{\theta}\,\bigr]^{ab}_{\alpha \beta}(x,y)\  
\phi\bigl[i,\,f,\,r,\,S,\,\vec{\theta},\,\tilde{\alpha},\,\tilde{t}\,\bigr]^b_{\beta}(y)\ =\ 
\bigl(S\,\eta[i]\bigr)^a_\alpha(x)\ \delta_{\alpha \tilde{\alpha}}\ \delta_{t_x\tilde{t}}\,\\
&\begin{split}
    \sum_{y} D\bigl[f_2,\,r_2,\,\vec{\theta}_2\bigr]^{ab}_{\alpha\beta}(x, y) \  
    \Phi\bigl[i,\,f_2,\,r_2,\,f_1,\,r_1,\,\Gamma_2,\,S_2,\,\vec{\theta}_2,\, \vec{\theta}_1,\,&\tilde{\alpha},\,
    \tilde{t},\,\tilde{t}+t_S\bigr]_{\beta}^b(y)\\  
    &=\  \Gamma_{2\,\alpha \beta}\  
    \bigl(S_2\,\phi\bigl[i,\,f_1,\,r_1,\,\vec{\theta}_1,\,\tilde{\alpha},\,\tilde{t}\,\bigr]\bigr)_{\beta}^a(x) 
    \ \delta_{t_S\, t_x - \tilde{t}}\,.
\end{split}    
\end{aligned}
\right.
\]
We compute the ``charged'' $B$ and $D$ two-point correlators 
$C^{(2)\,hl}_{\vec{\theta};S_1\,\Gamma_1;S_2\,\Gamma_2}(t)$ which read
\cite{Frezzotti:2009dr}:
\begin{eqnarray*}
C^{(2)\,hl}_{\vec{\theta};S_1\,\Gamma_1;S_2\,\Gamma_2}(t)&=&
\frac{1}{2}\sum_{r=\pm 1} \left\langle {\rm Tr} 
\sum_{\vec{x},\vec{y}} \Gamma_1\, 
{\mathscr S}^{S_1}_l(r;\,\vec{y},\,\tilde{t};\,\vec{x},\,\tilde{t}+t)\,\Gamma_2\,
{\mathscr S}^{S_2}_h(-r;\,\vec{x},\,\tilde{t}+t;\,\vec{y},\,\tilde{t})\right\rangle\,,\\
&=&
\frac{1}{2}\sum_{r=\pm 1}
\frac{1}{N} \sum_{n=1}^N \left\langle {\rm Tr} 
\left\{\sum_{\vec{x}} (\Gamma_1 
\gamma_5)_{\tilde{\alpha}\tilde{\beta}}\, 
\phi^*\bigl[n,\,l,\,r,\,S_1,\,\vec{0},\,\tilde{\beta},\,\tilde{t}\,\bigr]^{b}_{\alpha}(\vec{x},\tilde{t}+t)\right.\right.\\
&&\hspace{6cm}\Biggl.\Biggl.\times\,(\gamma_5 \Gamma_2)_{\alpha \beta} (S_2\, 
\phi\bigl[n,\,h,\,r,\,\vec{\theta},\,\tilde{\alpha},\,\tilde{t}\bigr])^b_{\beta}(\vec{x},\tilde{t}+t)\Biggr\}
\Biggr\rangle\,,
\end{eqnarray*}
where $\left\langle ... \right\rangle$ stands for the gauge ensemble 
average and $h\equiv c$ or $b$.\par
\begin{figure*}[htbp]
\begin{center}
\includegraphics*[width=5cm, height=3cm]{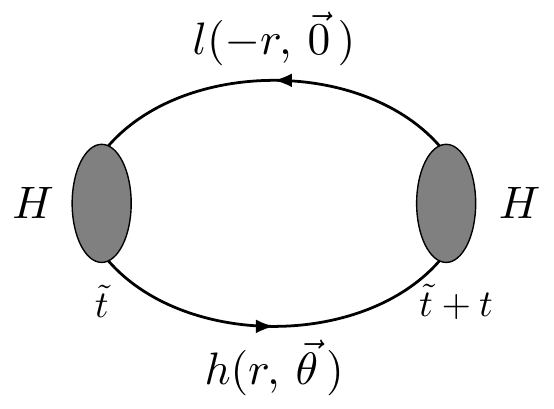}\quad\quad
\includegraphics*[width=5cm, height=3cm]{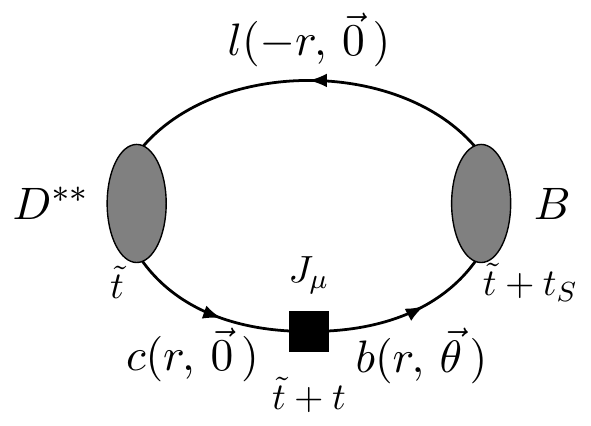}\\
\end{center}
\caption{\label{figcorrel}\em Kinematical configuration of the 2-pt correlators 
(left) and 3-pt correlators (right) we compute.}
\end{figure*}

We recall that, in Twisted-Mass QCD, quark 
propagators have the hermiticity property:
$${\mathscr S}_q(r;\,x;\,y)\ =\ \gamma_5\, 
{\mathscr S}^\dag_q(-r;\,y;\,x)\, \gamma_5\,.$$
We also compute the ``neutral" $B \to D$ three-point correlators 
$C^{(3)\,b\Gamma c}_{\vec{\theta};S_1\,\Gamma_1;\Gamma;S_2\,\Gamma_2}(t,t_s)$ 
which read:
\begin{eqnarray*}
C^{(3)\,b\Gamma c}_{\vec{\theta};S_1\,\Gamma_1;\Gamma;S_2\,\Gamma_2}(t,t_S)
&=&\frac{1}{2}\sum_{r=\pm 1}\left\langle {\rm Tr}\sum_{\vec{x},\vec{y},\vec{z}}
\Gamma\,{\mathscr S}_c(r,\,\vec{0};\,\vec{z},\,\tilde{t}+t;\,\vec{y},\,\tilde{t})\,\Gamma_1\,
{\mathscr S}^{S_1}_l(-r,\,\vec{0};\,\vec{y},\,\tilde{t};\,\vec{x},\,\tilde{t}+t_S)\right.\\
&&\hspace{8cm}\Biggl.\times\Gamma_2\,
{\mathscr S}^{S_2}_b(r,\,\vec{\theta};\,\vec{x},\,\tilde{t}+t_S;\,\vec{z},\,\tilde{t}+t)
\Biggr\rangle\,,\\
&=&\frac{1}{2}\sum_{r=\pm 1}
\frac{1}{N} \sum_{n=1}^N \left\langle {\rm Tr} 
\left\{\sum_{\vec{x}} (\Gamma 
\gamma_5)_{\tilde{\alpha}\tilde{\beta}} \,
(\phi\bigl[n,\,c,\,r,\,S_1,\,\vec{0},\,\tilde{\beta},\,\tilde{t}\,\bigr])^{b}_{\alpha}(\vec{x},\,\tilde{t}+t)\right.\right.\\
&&\hspace{4cm}\Biggl.\Biggl.\times\, 
(\gamma_5 \Gamma_1)_{\alpha \beta} \,
\Phi^*\bigl[n,\,b,\,-r,\,l,\,r,\,S_2,\,\Gamma_2,\,\vec{\theta},\,\vec{0},\,\tilde{\alpha},\,\tilde{t},\,\tilde{t}+t_S\bigr]^{b}_{\beta}
(\vec{x},\,\tilde{t}+t)\Biggr\}\Biggr\rangle\,.
\end{eqnarray*}
Those two types of correlators are depicted in Figure \ref{figcorrel}. On each of the two
ensembles, we estimate the statistical error from a jackknife procedure.

\section{Masses and energies\label{secspectro}}

\begin{figure*}[t!]
\begin{center}
\includegraphics*[width=7cm, height=5cm]{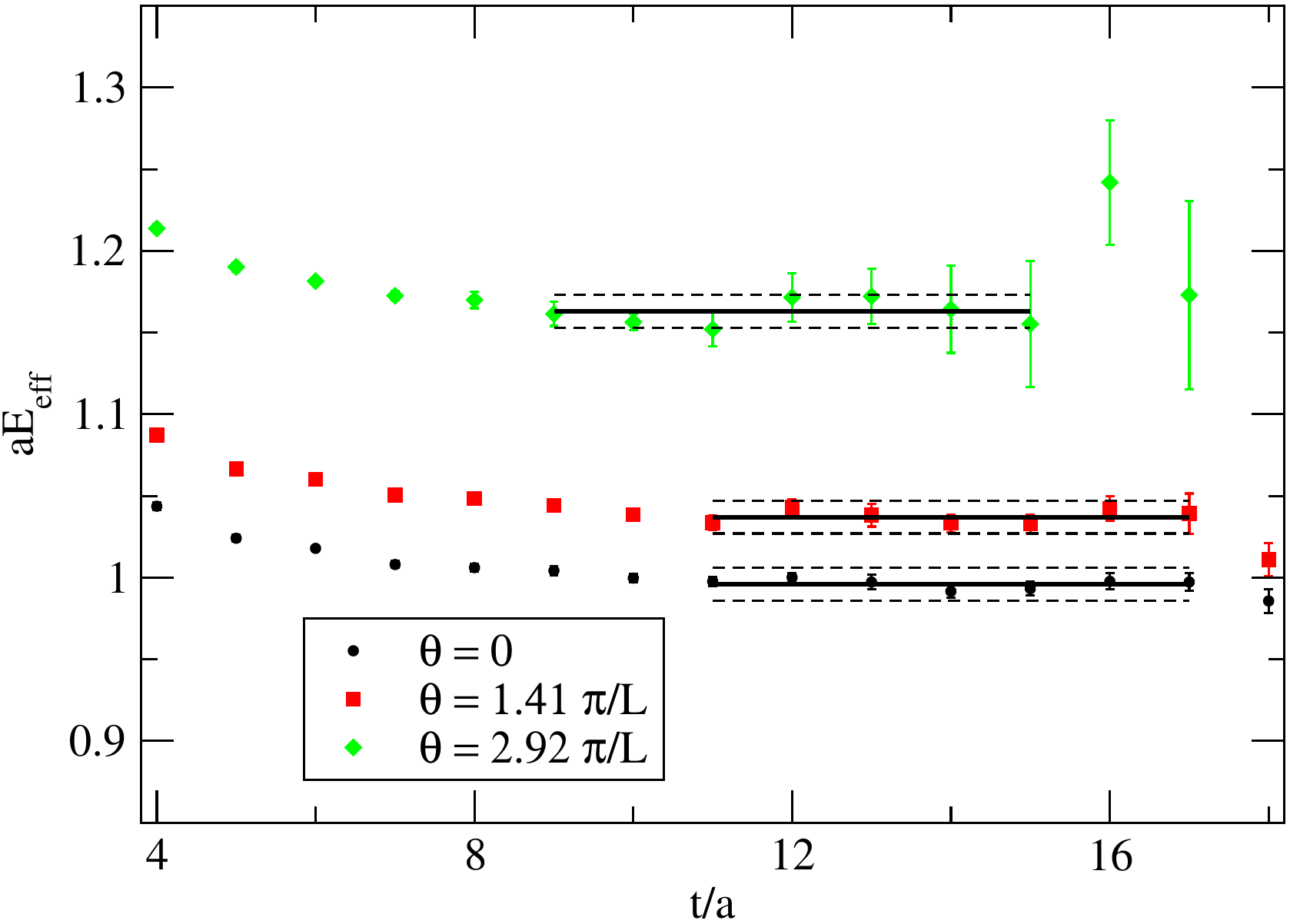}\quad\quad
\includegraphics*[width=7cm, height=5cm]{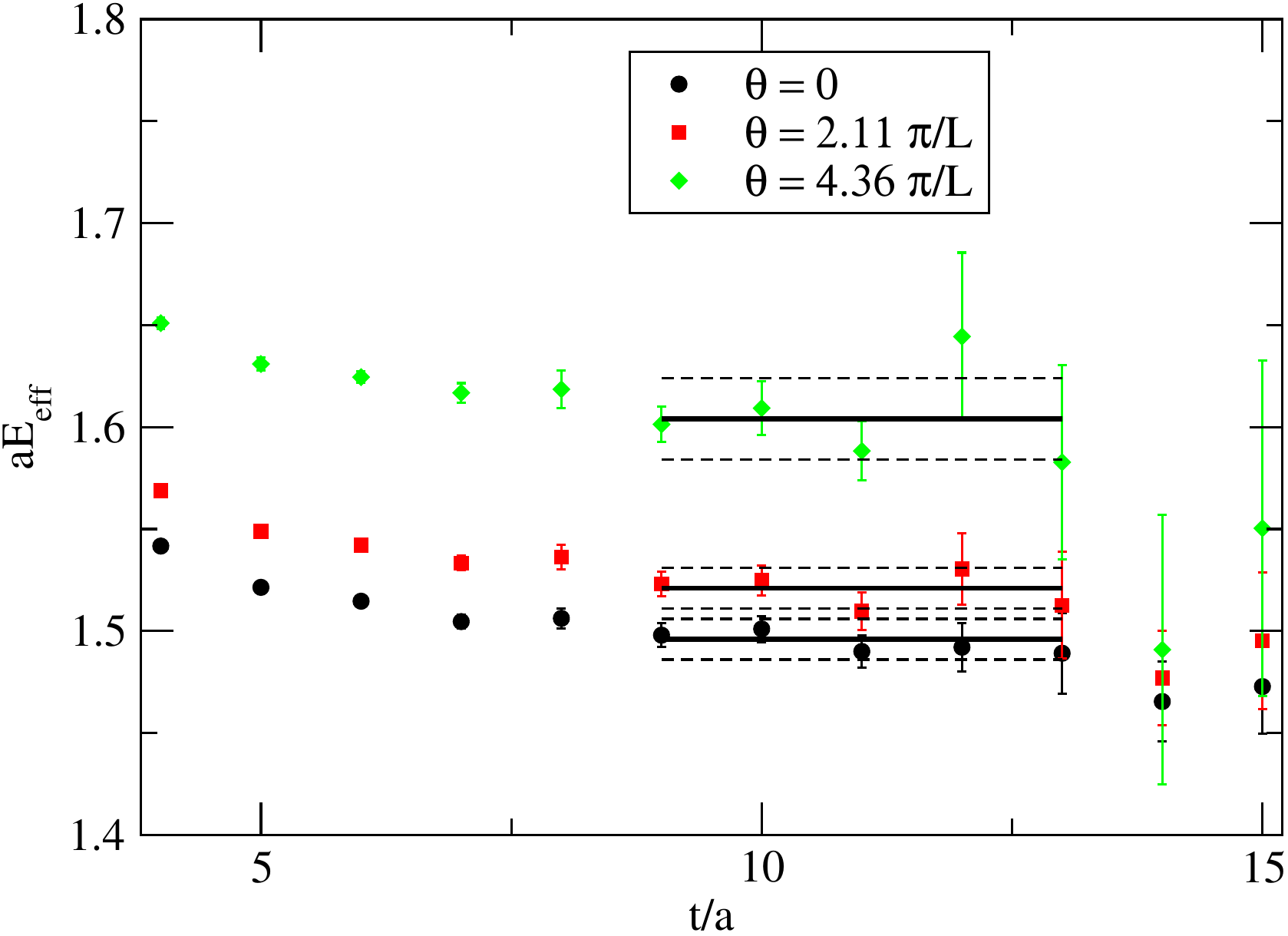}\\
\end{center}
\caption{\label{figmassPS} \em Effective energies of "$B$"-mesons measured with the
ETMC ensemble ($\beta=3.9$, $\mu_{\rm sea}=\mu_l=0.0085$):
$\mu_h=0.3498$ (left) and $\mu_h=0.6694$ (right).}
\end{figure*}

\begin{figure*}[t!]
\begin{center}
\includegraphics*[width=7cm, height=5cm]{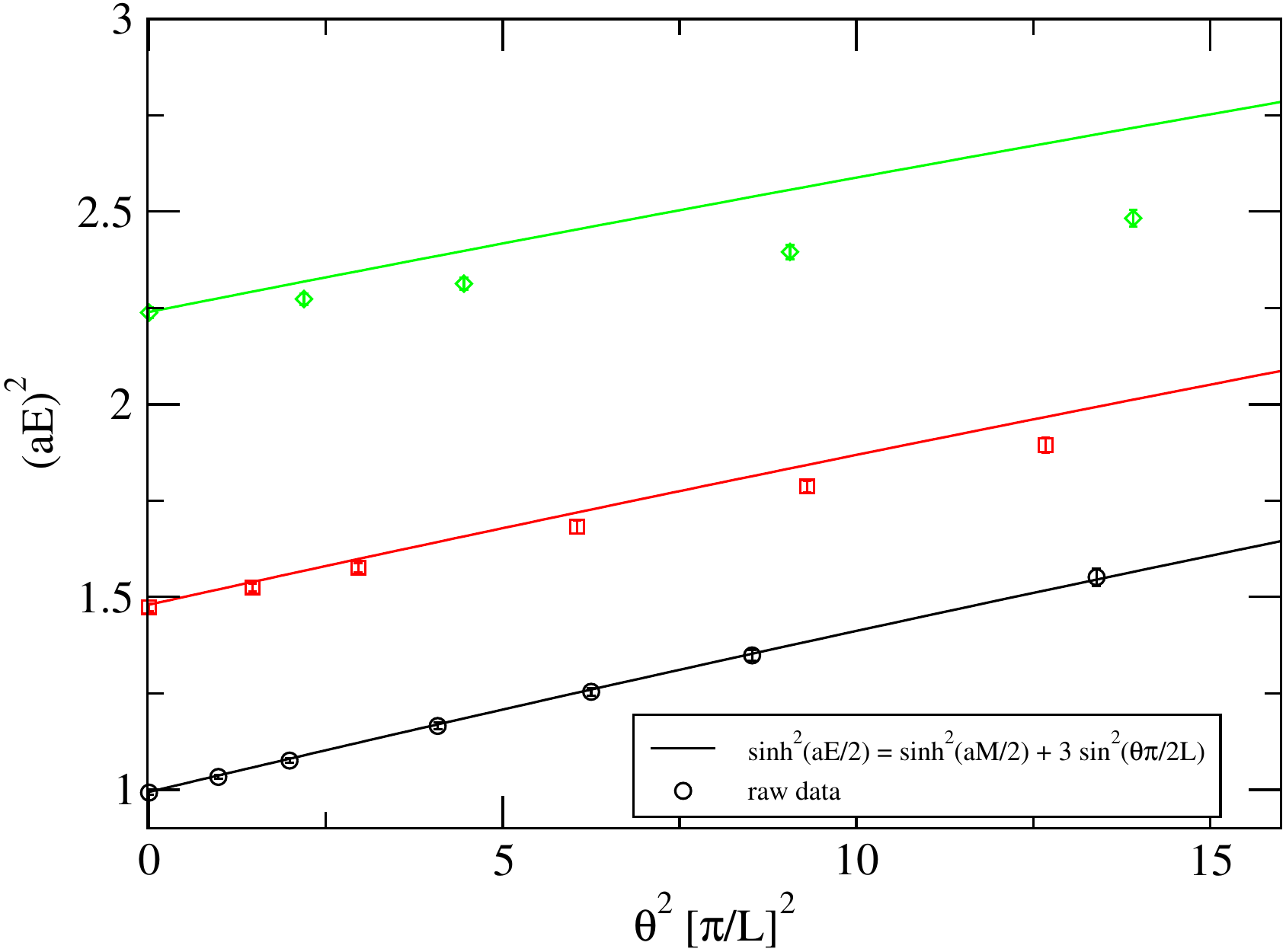}\\
\end{center}
\caption{\label{figenergymomentum} \em Comparison of the "$B$"-mesons energies with the
dispersion relation, at the ETMC ensemble ($\beta=3.9$, $\mu_{\rm sea}=\mu_l=0.0085$). 
The energy have been rescaled to 1 for the lightest $B$ at rest. The black, red 
and green points correspond to the three $B$ masses in increasing order.}
\end{figure*}

We decide to concentrate our effort on the analysis of smeared-smeared 2-pt correlators 
because the benefit of such a technique has been already clearly observed in a previous work 
by ETMC \cite{CarrascoDD}. Masses and energies of pseudoscalar $B$ and $D$ mesons are first 
extracted from a fit of the form
\begin{equation}\nonumber
C_{PP}(t,\vec{\theta})=\frac{{\cal Z}^{2\,(1)}}{2 E^{(1)}_P(\vec{\theta})} 
\left( e^{-E^{(1)}(\vec{\theta}) t} + e^{-E^{(1)}(\vec{\theta}) (T-t)}\right),
\;\; {\cal Z}^{2\,(1)}=\brakket{H^{(1)}}{O^{H\dag}_P}{0}
\end{equation}
in a time range where the contribution from the first excitation is small compared to the
statistical error. The stability of the fit is checked by enlarging the time interval and
adding a second exponential in the fomula, i.e.: 
\begin{equation}\nonumber
C_{PP}(t,\vec{\theta})=\sum_{i=1}^2 \frac{{\cal Z}^{2\,(i)}}{2 E^{(i)}_P(\vec{\theta})} 
\left( e^{-E^{(i)}(\vec{\theta}) t} + e^{-E^{(i)}(\vec{\theta}) (T-t)}\right)
\end{equation}
The last step in the analysis is to measure the effective energy $E_P \equiv E^{(1)}_P$ of the 
ground state from the ratio
\begin{equation}\nonumber
\frac{C_{PP}(t+1,\vec{\theta})}{C_{PP}(t,\vec{\theta})}=
\cosh(E_P(\vec{\theta})) + \sinh(E_P(\vec{\theta}))
\tanh [E_P(\vec{\theta}) (t-T/2)]
\end{equation}
We show in Figure \ref{figmassPS} examples of plateaus for "$B$"-mesons energies
at three different momenta.
We study the dispersion relation to get an idea of the magnitude of cut-off effects.
We display in Figure \ref{figenergymomentum} the $B$ meson energies and compare them to the theoretical
formula
\begin{equation}
\sinh^2 [aE(\theta)/2] = \sinh^2 [aM/2] + 3 \sin^2(\theta/2)\qquad\text{where} \qquad M\equiv E(0)
\end{equation}
The agreement is good at the two lightest heavy masses but really bad at the heaviest one:
cut-off effects are pretty large.\par
\begin{figure*}[htb]
\begin{center}
\includegraphics*[width=7cm, height=5cm]{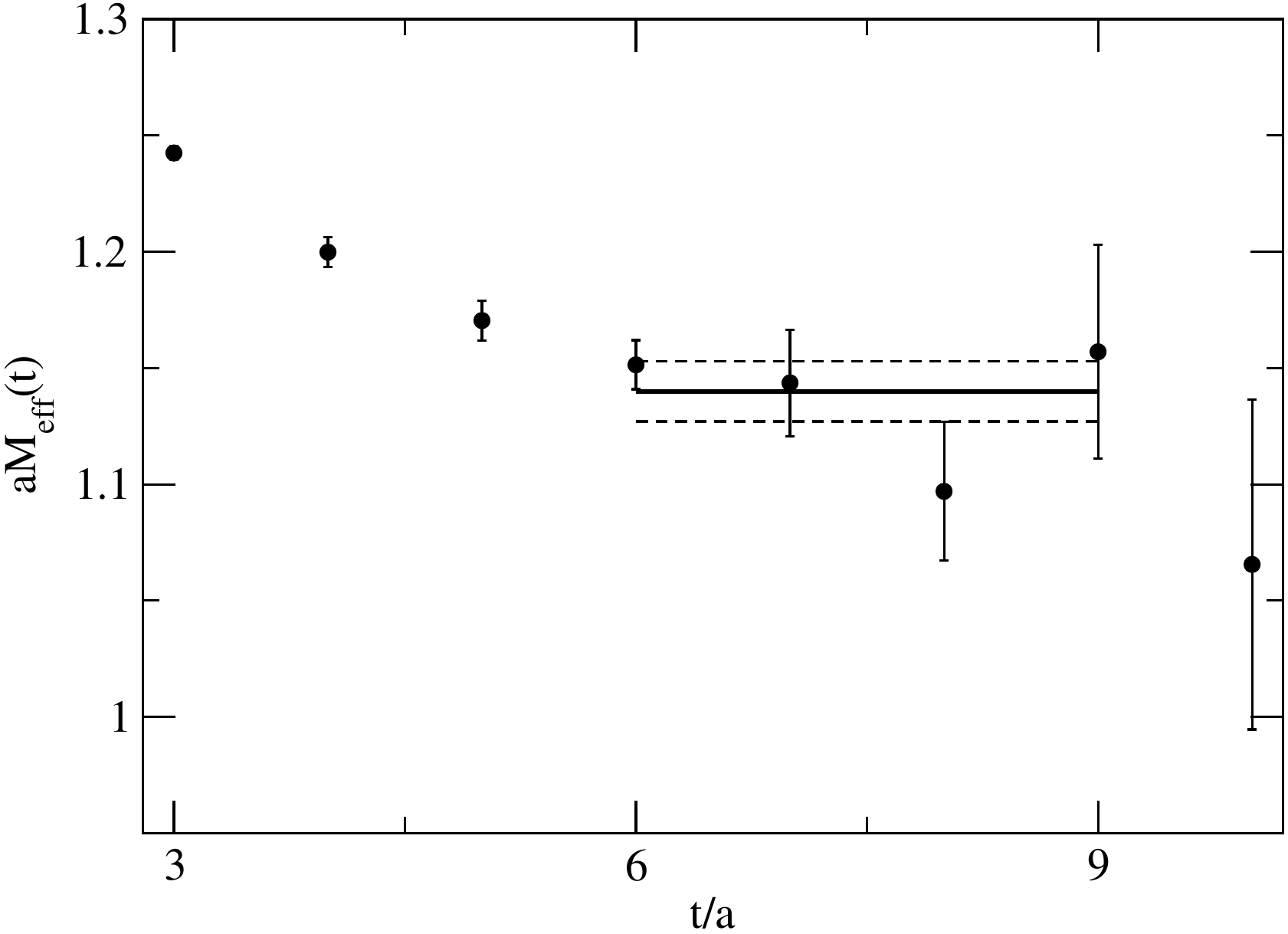}\quad\quad
\includegraphics*[width=7cm, height=5cm]{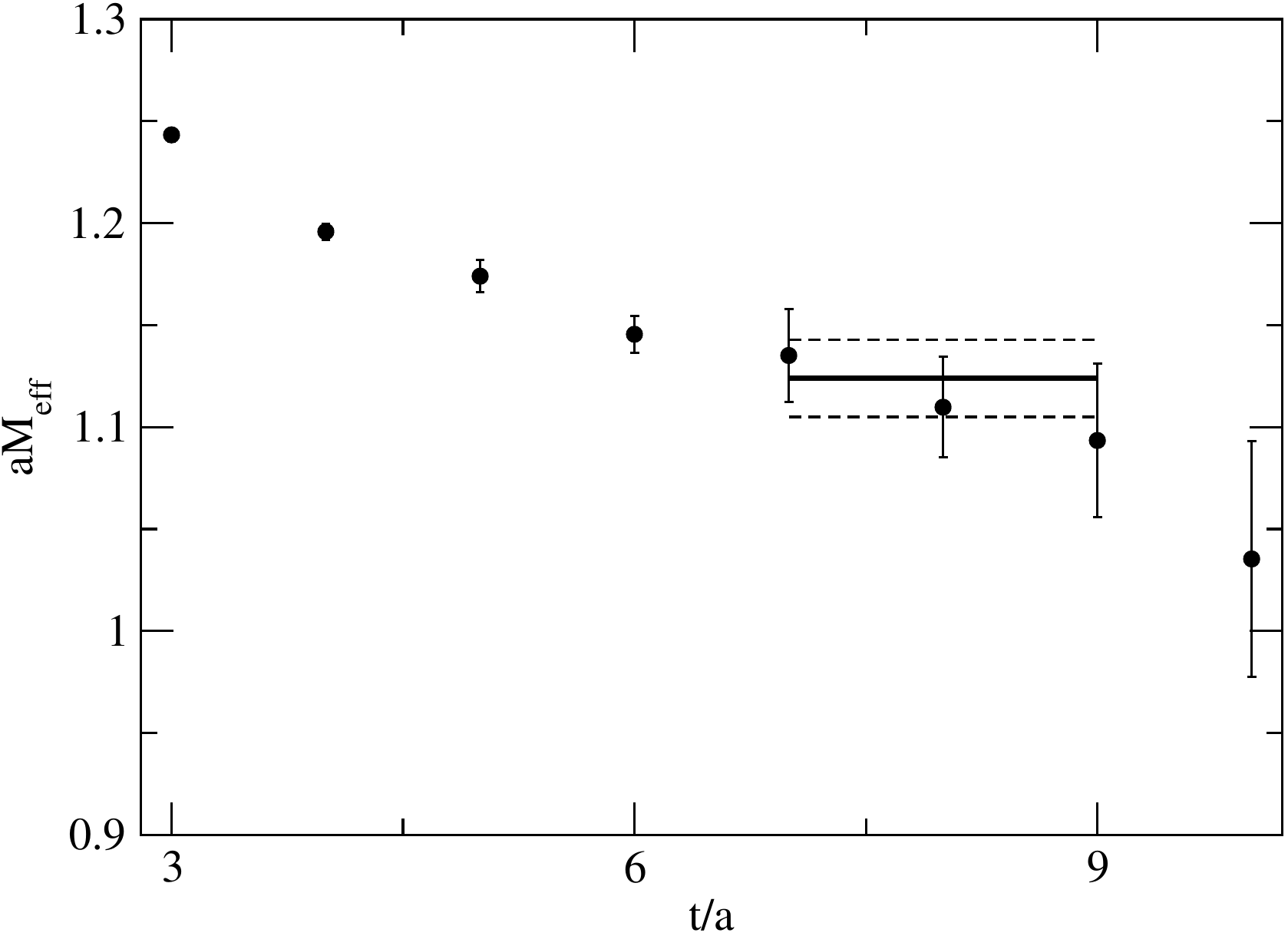}\\
\end{center}
\caption{\label{figmass2p}\em Effective mass of the $D^*_2$ meson measured with the
ETMC ensemble ($\beta=3.9$, $\mu_{\rm sea}=\mu_l=0.0085$):
$\frac{C^{(2)}_{2^+,E}(t)}{C^{(2)}_{2^+,E}(t+1)}$ (left) and 
$\frac{C^{(2)}_{2^+,T_2}(t)}{C^{(2)}_{2^+,T_2}(t+1)}$ (right).}
\end{figure*}

\noindent
Interpolating fields of the $2^+$ state are given by the formula 
$O^{(\lambda)}=\epsilon^{*(\lambda)}_{\mu\nu} \bar{\chi}_c \gamma_\mu \nabla_\nu \chi_l$, 
$\lambda=\pm 2,\, \pm 1,\, 0$. Actually we choose to use linear combinations of those
interpolating fields that read:
\[
\left\{
\begin{aligned}
\tilde{O}^{(1)}&\ =\ \frac{1}{\sqrt{2}}(O^{(+2)}+O^{(-2)})\ =\ \frac{1}{\sqrt{2}} \bar{\chi}_c (\gamma_1 \nabla_1 -\gamma_2 \nabla_2) 
\chi_l\\[3mm]
\tilde{O}^{(2)}&\ =O^{(0)}\ =\ -\frac{1}{\sqrt{6}} \bar{\chi}_c (\gamma_1 \nabla_1 + \gamma_2 \nabla_2
- 2 \gamma_3 \nabla_3) \chi_l\\[3mm]
\tilde{O}^{(3)}&\ =\ \frac{1}{\sqrt{2}}(O^{(+2)}-O^{(-2)})\ =\ -\frac{i}{\sqrt{2}}\bar{\chi}_c (\gamma_1 \nabla_2 + \gamma_2 \nabla_1) \chi_l\\[3mm]
\tilde{O}^{(4)}&\ =\ \frac{1}{\sqrt{2}}(O^{(+1)}+O^{(-1)})\ =\ \frac{i}{\sqrt{2}}\bar{\chi}_c (\gamma_2 \nabla_3 + \gamma_3 \nabla_2) \chi_l\\[3mm]
\tilde{O}^{(5)}&\ =\ \frac{1}{\sqrt{2}}(O^{(+1)}-O^{(-1)})\ =\ -\frac{1}{\sqrt{2}}
\bar{\chi}_c (\gamma_1 \nabla_3 + \gamma_3 \nabla_1) \chi_l
\end{aligned}
\right.
\]
The two first interpolating fields live in the $E$ representation of the $O_h$ cubic group 
symmetry of rotations and inversion in a 3-d spatial lattice, while
the three last live in the $T_2$ representation of that group \cite{LacockVY}.
We finally consider the $r$-symmetrized smeared-smeared 2-pt correlators
\begin{equation}
C^{(2)}_{2^+,E}(t)=\frac{1}{2} \left[\langle \sum_{\vec{x},\vec{y}} 
\tilde{O}^{(1)}_S(\vec{y},t+\tilde{t})\tilde{O}^{\dag(1)}_S(\vec{x},\tilde{t})
\rangle + \langle \sum_{\vec{x},\vec{y}} 
\tilde{O}^{(2)}_S(\vec{y},t+\tilde{t})\tilde{O}^{\dag(2)}_S(\vec{x},\tilde{t})
\rangle\right]
\end{equation} and 
\begin{equation}
C^{(2)}_{2^+,T_2}(t)=\frac{1}{3} \left[\langle \sum_{\vec{x},\vec{y}} 
\tilde{O}^{(3)}_S(\vec{y},t+\tilde{t})\tilde{O}^{\dag(3)}_S(\vec{x},\tilde{t})
\rangle + \langle \sum_{\vec{x},\vec{y}} 
\tilde{O}^{(4)}_S(\vec{y},t+\tilde{t})\tilde{O}^{\dag(4)}_S(\vec{x},\tilde{t})
\rangle + \rangle + \langle \sum_{\vec{x},\vec{y}} 
\tilde{O}^{(5)}_S(\vec{y},t+\tilde{t})\tilde{O}^{\dag(5)}_S(\vec{x},\tilde{t})
\rangle\right]
\end{equation} 
The masses we extract by studying the ratios 
$\frac{C^{(2)}_{2^+,E}(t)}{C^{(2)}_{2^+,E}(t+1)}$ and 
$\frac{C^{(2)}_{2^+,T_2}(t)}{C^{(2)}_{2^+,T_2}(t+1)}$ are in principal equal: any 
discrepancy comes from cut-off effects. We show in Figure \ref{figmass2p} that, 
indeed, lattice artefacts are present.

\begin{figure*}[t!]
\begin{center}
\includegraphics*[width=7cm, height=5cm]{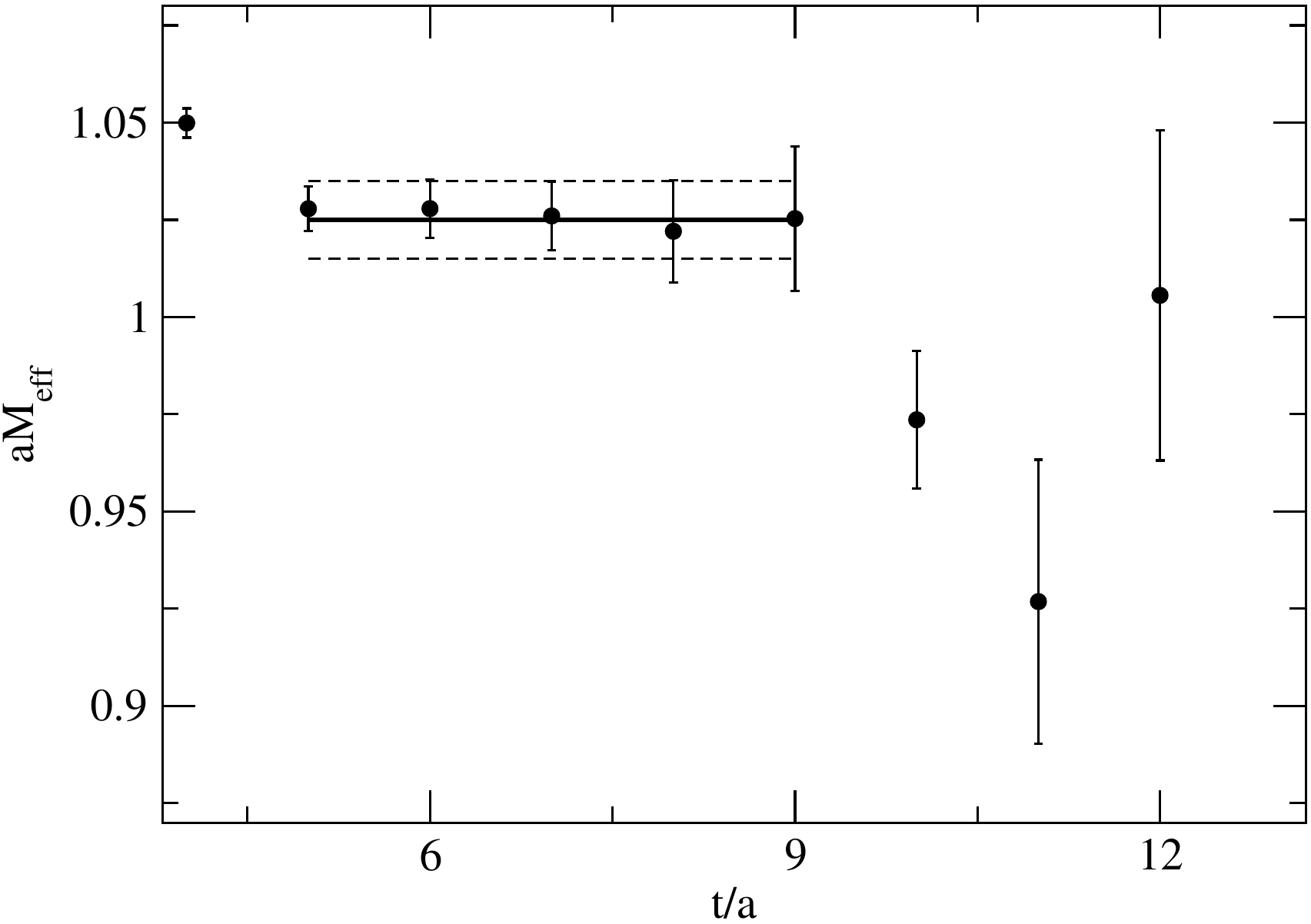}
\end{center}
\caption{\label{figmass0p}\em Effective mass of the $D^*_0$ meson measured with the
ETMC ensemble ($\beta=3.9$, $\mu_{\rm sea}=\mu_l=0.0085$).}
\end{figure*}

As parity is broken by Twisted-mass action at finite lattice spacing \cite{Frezzotti:2003ni}
and the states we consider are not made with quarks of the same flavour doublet, contrary 
to what is discussed in section 5.2 of \cite{Shindler:2007vp}, the scalar $D$ meson 
can in principle mix
with the pseudoscalar $D$ meson. We have to build a matrix of correlators $\{C_{ij}(t)\}$ and solve 
a Generalised 
Eigenvalue
Problem (GEVP) \cite{MichaelNE} - \cite{BlossierKD}. We study a $2\times 2$ system whose
entries correspond to the interpolating fields with Dirac structures 
$\bar{\chi}_c \gamma^5 \chi_l$ and 
$\bar{\chi}_c \chi_l$:

\begin{equation}\nonumber 
C_{ij}(t)=\left [ \begin{array}{ll} C^{(2)}_{ \bar{\chi}_c \gamma_5 \chi_l;\bar{\chi}_l \gamma_5 \chi_c}
(t) 
&C^{(2)}_{\bar{\chi}_c \gamma_5 \chi_l;i\bar{\chi}_l \chi_c}(t)\\[2mm]
C^{(2)}_{-i\bar{\chi}_c \chi_l;\bar{\chi}_l \gamma_5 \chi_c}(t) & 
C^{(2)}_{\bar{\chi}_c \chi_l;\bar{\chi}_l \chi_c}(t) 
\end{array} \right ]
\end{equation}
We solve the system
\begin{equation}\label{eq:cij}
C_{ij}(t) v^{(n)}_j(t,t_0)=\lambda^{(n)}(t,t_0) C_{ij}(t_0) v^{(n)}_j(t,t_0)
\end{equation}
We set $t_0=3$ $(\beta=3.9)$ and 5 $(\beta=4.05)$. $\lambda^{(n)}(t,t_0)$ and 
$v^{(n)}(t,t_0)$ are the eigenvalues and eigenvectors of the matrix $C^{-1}(t_0)C(t)$.
The effective mass $m_{D^*_0}$ of the scalar meson is given by:
\begin{equation}
\lambda^{(2)}(t,t_0)=\frac{\cosh[m_{D^*_0}(T/2-t)]}{\cosh[m_{D^*_0}(T/2-t_0)]}
\end{equation}
We show in Figure \ref{figmass0p} $m_{D^*_0}$ for the ensemble
($\beta=3.9, \mu_{\rm sea}=0.0085$). The signal is unfortunately quite short, but still acceptable
for our qualitative study.

We collect in the Appendix all the masses and energies that we extracted in our analysis. The total 
error includes the statistical one and the discrepancy of results when we change the time range
$[t_{\rm min}, t_{\rm max}]$ of fits by $t_{\rm min}\pm 1$ and $t_{\rm max}\pm 1$, when we take 
different $t_0$ in the range [3, 6] and, in the case of pseudoscalar $B$ mesons, when we perform a 
2-states exponential fit.
\subsection{$\boldsymbol{M_{D^\ast_2}-M_D}$ and $\boldsymbol{M_{D^\ast_0}-M_D}$ in the continuum limit and
experiment} In Table~\ref{tabmassenergies} in the appendix are given the masses of the ${D^\ast_2},
{D^\ast_0}$ and $D$ mesons. It is interesting to perform an extrapolation  to
the continuum and compare with the experimental data. For the latter we will
take the $c \bar s$ mesons.  It does not change anything for the tensor meson as
compared to the non-strange charmed mesons  ($M_{D^\ast_2}(2460) -M_D \simeq 
M_{D^\ast_{2s}}(2573) -M_{D_s}$) but for the scalar mesons it does, since the
$D^\ast_{0s}(2317)$ has a narrow and clear signal which is not the case of the
$D^\ast_{0}(2400)$ whose signal is very broad due to its S-wave  decay into
$D \pi$. 

\begin{table}[htb]
\begin{center}
\begin{tabular}{||c|c|c|c||}
\hhline{|t:====:t|}
lattice spacing (fm) &inverse spacing (GeV)&  scalar & tensor \\
\hhline{|:=:=:=:=|}
0.085(3)  & 2.32(8) & 0.58(9) &0.88(7)  \\
0.069(2) & 2.85(8) & 0.51(5) & 0.88(5)\\
0.0054(2) & 3.65(8) & 0.51(5) &        \\
0.0    & $\inf$& 0.44(7)& 0.88(7) \tiny{$\begin{pmatrix} +15 \\ -18
\end{pmatrix}$} \\
Exp.  & & 0.349 & 0.605 \\ 
\hhline{|b:====:b|} 
\end{tabular}
\end{center}
\caption{\em Mass differences $M_{D^\ast_0}-M_D$ (third column) and 
$M_{D^\ast_2}-M_D$ (fourth column) for $\beta=3.9$ (first line), $\beta=4.05$
(second line) and $\beta=4.2$ (third line), the latter being  
for the moment restricted to $M_{D^\ast_0}-M_D$.
 We use the lattice spacings reported in Table~\ref{tabsimul}. 
The extrapolation to vanishing lattice spacing is in the fourth line and the
experimental data $M_{D^\ast_{0s}}(2317) -M_{D_s}$ and $M_{D^\ast_{2s}}(2573)
-M_{D_s}$ are in the fifth line. 
The errors combine in quadrature the errors on the masses in lattice units
and on the lattice spacing. The fitting windows are reported in
 table~\ref{tabmassenergies}. The
dependence of the value extrapolated to continuum strongly depends on  
the lattice spacing. Our uncertainty does not incorporate the systematic
one due to the choice of the fitting windows.
Our mass differences extrapolated to the continuum are above experiment.
}
\label{tab:spectro}
\end{table}

We perform the extrapolation to the continuum limit and compute the error using
jackknife. We show our results in Table~\ref{tab:spectro}. We see that the
agreement with experiment is not so good but the errors are large. 
For the moment we cannot say much, except that the future will tell
 whether this issue comes mainly from statistical fluctuations or whether the 
lattice regularization is in cause. A similar conclusion was made 
recently in a lattice study with 
${\rm N_f}=2+1+1$ dynamical quarks regularised by TmQCD \cite{Wagner:2013laa}.
Also, we have not considered yet the possible effect of the
opening of the decay channel $D^*_0 \to D\pi$ $S$-wave state as proposed
in \cite{Mohler:2012NA}.

\section{$\boldsymbol{B}$ decay to the scalar $\boldsymbol{D^*_0}$ charmed meson}

In this section we will restrict to the zero recoil kinematics, in other words the 
initial $B$ meson is taken at rest. We restrict ourselves to this simpler
case because the momentum dependance of the $B \to D^*_0$ decay is very difficult 
to study on the lattice (3-pt correlators are very noisy) and
we cannot yet say anything significant about it, but also because the non vanishing
of the zero recoil amplitude is of utmost phenomenological relevance : in the 
infinite mass limit, the $B \to D^*_0$ amplitude vanishes at zero recoil~\cite{ISGW2}.
This forbids the decay into a S-wave between the lepton 
pair and the $D^*_0$  since an S-wave clearly does not vanish at zero
recoil. The S-wave is a major contribution to these decays since their
available phase space is rather small, and higher waves are suppresed by the
so-called centrifugal barreer effect. With finite heavy quark masses we will
show that the zero recoil amplitude {\it does not vanish}, there is a non
vanishing S-wave and this may change
drastically the ratio between  $\Gamma(B \to D^*_2)$ and $\Gamma(B \to D^*_0)$ since the $B \to
D^*_2$ decay amplitude does vanish at zero recoil whether the mass of the $c$
and $b$ is taken infinite or finite, thus implying only a D-wave decay. The
possible importance of a non vanishing zero recoil amplitude was stressed
in~\cite{Leibovich:1997em} where the authors estimated subleading corrections to
the infinite mass limit : although subleading in the $\Lambda/m_{c,b}$ expansion, 
the S-wave may not be negligible. 
Our computation confirms this conclusion as shall be seen. 
\subsection{Computation of the amplitude ratio $\boldsymbol{(B \to D^*_0})$ over 
$\boldsymbol{(B \to D)}$ at zero recoil}  
We will compute the ratio of amplitudes $A(B \to D^*_0)/A(B \to D)$. We take
 $A(B \to D)$ as a benchmark since it is experimentally fairly well known, 
and $D$ being a $J=0$  state as $D^*_0$, it is expected that the momentum 
dependence of these decays will be rather similar. 
 We recall some formulae
neglecting for the moment the $D-D^*_0$ mixing due to the parity violation at 
finite lattice spacing when using twisted mass quarks.
The matrix elements $\langle B|V_{0}|D\rangle$ and $\langle B|A_0|D^*_0\rangle$
are given by the following ratio
\beq\label{c3surc2}
\langle B|O|H_c\rangle = \frac{C^{(3)}_{\rm  BOH_c}(t_{p},t,t_s) \sqrt{Z_{H_c}/Z_B}Z_O}
{C^{(2)}_{H_c}(t-t_s)\exp(-E_B\,(t_{p}-t))/(2E_B)}\,,
\eeq
where $t_s< t < t_p$ are respectively the source, current and sink times 
(cf Fig \ref{figcorrel}). 

$Z_0 = Z_V (Z_A)$ when $H_c = D (D^*_0)$. $Z_{H_c}$ and $Z_B$ are defined from the
fit of the 2-pt correlators of the $H_{c}$ and $B$ mesons, assuming we are far enough from the center of the lattice to be allowed to neglect
the backward exponential in time while the contribution of excited states is small.   
\beq \label{deuxpoints}
C^{(2)}_B(t,\vec{\theta})= \frac{Z_B}{2 E_B(\vec{\theta})} \exp(-E_B (\vec{\theta}))t; \qquad 
C^{(2)}_{H_c}(t,0)=\frac {Z_{H_c}}{2 M_{H_c}} \exp\left(-M_{H_c} t\right)\,.
\eeq
Then we compute
\beq\label{ratioDscatoD}
\frac {\langle B|A_0|D^*_0\rangle}  {\langle B |V_0|D\rangle} =
\frac {C^{(3)}_{\rm BA_0D^*_0}(t_p,t,t_s)C^{(2)}_{D}(t-t_s)Z_A\sqrt{Z_{D^*_0}}} 
 {C^{(3)}_{\rm BV_0D}(t_{p},t,t_s)C^{(2)}_{D^*_0}(t-t_s)Z_V\sqrt{Z_D}}\,.
\eeq
\begin{table}[htb]
\begin{center}
\begin{tabular}{||c|c||}
\hhline{|t:==:t|}
$v^{(1)}$ & $v^{(2)} $\\
\hhline{|:=:=:|}
0.97 & 0.28\, i \\
0.22\ i & 0.96 \\
\hhline{|b:==:b|} 
\end{tabular}
\end{center}
\caption{\em Values of the approximately orthonormalised eigenvectors $v^{(1)}$ and $v^{(2)}$, eq.~(4.5),
for $\beta=3.9$, $t_0 = t_s = 3$, $t = 7$ and $t_p = 14$. $v^{(1)}v^{(2)\dagger} = 0.07\,i $, rather small as expected in eq~\ref{eq:ortho}. }
\label{tab:v1v2}
\end{table}
\subsection{Taking into account the scalar-pseudoscalar mixing}
In section \ref{secspectro} we have detailed the Generalised Eigenvalue method. 
As explained there and in section~\ref{sec:simulation}, we restrict ourselves to a $2 \times 2$ matrix of smeared and stochastic 2-pt correlators.
The largest (smallest) eigenvalue $\lambda^{(1)}$ ($\lambda^{(2)}$ ) will be related 
to the mass of the $D$ ($D^*_0$) state. The corresponding eigenvectors 
give the linear combination of
 $\bar{\chi}_{c} \gamma_5 \chi_{l}$ and $\bar{\chi}_{c}\chi_{l}$ interpolating fields
 that have the largest coupling to the $D(D^{*}_{0})$ state. 
The eigenvectors turn out to be not far from orthogonal
\beq\label{eq:ortho}
v^{(1)}v^{(2)\dagger} = \sum_{k=1,2} v^{(1)}_k v^{(2)\star}_k = 0.07 i \simeq 0\,,
\eeq
 as seen for example in table~\ref{tab:v1v2}. One might say that 0.07 is not so small but 
this is not surprising, since there are other states in which the B might decay than 
only the ground state scalar and pseudoscalar that we consider in our analysis. 
We can thus to a fair approximation orthonormalise the eigenvectors so that
\beq
v^{(i)}v^{(j)\dagger} \simeq \delta_{i,j}\,,
\eeq
without changing the eigenvalues, since in eq~(\ref{eq:cij})
the same factor multiplies both sides of the equation. 
Thus, we define the 2-pt correlator
 of $D^{*}_{0}$ meson by
\beq\label{eq:C2}
 \lambda^{(2)}(t-t_s,t_0)\sum_{i,j=1}^{2}v^{(2)\,\dagger}_{i}(t-t_s,t_{0})C^{(2)}_{ij}(t_0)v^{(2)}_{j}(t-t_s,t_{0}) &=& \sum_{i,j=1}^{2}v^{(2)\,\dagger}_{i}(t-t_s,t_{0})  C^{(2)}_{ij}(t-t_s) 
v^{(2)}_{j}(t-t_s,t_{0})\,, \nonumber \\
  \lambda^{(1)}(t-t_s,t_0)\sum_{i,j=1}^{2}v^{(1)\,\dagger}_{i}(t-t_s,t_{0})C^{(2)}_{ij}(t_0)v^{(1)}_{j}(t-t_s,t_{0}) &=& \sum_{i,j=1}^{2}v^{(1)\,\dagger}_{i}(t-t_s,t_{0})  C^{(2)}_{ij}(t-t_s) 
v^{(1)}_{j}(t-t_s,t_{0})\,,
\eeq
where we fit with 
\beq\label{eq:lambda}
 \lambda^{(2)}(t-t_s,t_0)&=& \exp\left(-M_{D^{*}_{0}} (t-t_s-t_0)\right) \quad
\lambda^{(1)}(t-t_s,t_0) =   \exp\left(-M_{D} (t-t_s-t_0)\right)\\ \nonumber
\eeq
since $\lambda^{(i)}(t_0,t_0) = 1$ from \eq{eq:C2}.
We define $Z_2^{(1)}$ and $Z_2^{(2)}$ from
\beq\label{eq:defZi}
v^{(2)\,\dagger}_{i}(t-t_s,t_{0})C^{(2)}_{ij}(t-t_s)v^{(2)}_{j}(t-t_s,t_{0})
&=& \frac {Z_2^{(2)}}{2 M_{D^{*}_0}} \exp\left(-M_{D^{*}_{0}} (t-t_s)\right )\,,\\   \nonumber
v^{(1)\,\dagger}_{i}(t-t_s,t_{0})C^{(2)}_{ij}(t-t_s)v^{(1)}_{j}(t-t_s,t_{0})
&=& \frac {Z_1^{(1)}}{2 M_{D^{*}_0}} \exp\left(-M_{D} (t-t_s)\right )\,,
\eeq
In \eq{c3surc2} we see that the factors $Z_i$ appear only via their square root.
\subsubsection{Symmetry properties of the matrix elements}
In the continuum it is obvious by parity conservation that 
\beq\label{eq:selection}
 \langle B|A_0|D \rangle = \langle B|V_0| D^{*}_{0}\rangle = 0
\eeq
 However 
parity is not conserved by the twisted mass quark action at finite lattice spacing. 
But this action has an exact symmetry~\cite{Frezzotti:2003ni}, the flavour-parity
\beq\label{R5}
{\mathscr R}^{\rm sp}_5 \equiv {\mathscr P} \otimes \left ( {\mu_l, \mu_c, \mu_b \leadsto 
-\mu_l, -\mu_c, -\mu_b} \right )
\eeq
where $\mathscr P$ is the spatial parity, and $\mu_l, \mu_c, \mu_b$ are the
twisted mass terms for the light, charm and beauty quarks. We assume we are
 at maximal twist (vanishing of $m_{\rm PCAC}$). Therefore if we 
use 
\beq\label{R5toC3}
C^{(3)\,\text{sym}}_{i,j,k}(t_p,t,t_s)\ \equiv\ \left ( 1 + 
{\mathscr R}^{\rm sp}_5 \right )\,C^{(3)}_{i,j,k}(t_p,t,t_s)
\eeq   
we get the validity of eq.~(\ref{eq:selection}) also for finite lattice spacing.
This symmetrisation will be assumed in the following.
\subsubsection{GEVP on the 3 pt correlators}
In this section we assume $t_s< t < t_p$ and $t_0 \le t-t_s$.
Starting from the 3-pt correlators $C^{(3)}_{B A_0(V_0) O_i} (t_{p},t,t_s)=\langle \bar{\chi}_b \gamma_5
\chi_l(t_p) A_0(V_0) (t) O_i(\bar{\chi}_c,\chi_l,t_s)\rangle$, $O_i(\bar{\chi}_c,\chi_l)\equiv 
\{\bar{\chi}_c \gamma_5 \chi_l,\,
\bar{\chi}_c \chi_l\}$ and following the authors \cite{BlossierMK} in their way of extracting the decay constant $f_B$,
we consider the projected 3-pt correlators 
\beq\label{troispoints}
\nonumber
C^{(3)\prime}_{B A_0 D^*_0} (t_{p},t,t_s)&=&\frac{\langle \bar{\chi}_b \gamma_5
\chi_l(t_p) A_0 (t) \bar{\chi}_c \chi_l(t_s) v^{(2)}_{\bar{\chi}_c \chi_l}(t-t_s,t_0)\rangle}
{\sqrt{v^{\dag(2)}_{i}(t-t_s,t_0) C^{(2)}_{ij}(t-t_s)v^{(2)}_j(t-t_s,t_0)}}\\
\nonumber
&\times&\left(\frac{\lambda^{(2)}(t_0+a,t_0)}{\lambda^{(2)}(t_0+2a,t_0)}\right)^{(t-t_s)/2a}
\frac{2E_B e^{(t_p - t)E_B}}{\sqrt{Z_B}}\sqrt{2m_{D^*_0}},\\
\nonumber
C^{(3)\prime}_{B V_0 D}(t_{p},t,t_s)&=&
\frac{\langle \bar{\chi}_b \gamma_5
\chi_l(t_p) V_0 (t) \bar{\chi}_c \gamma_5 \chi_l(t_s)v^{(1)}_{\bar{\chi}_c \gamma_5 \chi_l}(t-t_s,t_0)\rangle}
{\sqrt{v^{\dag(1)}_{i}(t-t_s,t_0) C^{(2)}_{ij}(t-t_s)v^{(1)}_j(t-t_s,t_0)}}\\
&\times&
\left(\frac{\lambda^{(1)}(t_0+a,t_0)}{\lambda^{(1)}(t_0+2a,t_0)}\right)^{(t-t_s)/2a}
\frac{2E_B e^{(t_p - t)E_B}}{\sqrt{Z_B}}\sqrt{2m_{D}}.
\eeq
We remind that the normalisation factor $Z^{(2)}_2$ cancels between $\langle \bar{\chi}_b \gamma_5
\chi_l(t_p) A_0(V_0) (t) \bar{\chi}_c (\gamma_5) \chi_l(t_s)\rangle$ and\\ 
$\sqrt{v^{\dag(2(1))}_{i}(t-t_s,t_0) C^{(2)}_{ij}(t-t_s)v^{(2(1))}_j(t-t_s,t_0)}$ while the factor $\left(\frac{\lambda^{(2(1))}(t_0+a,t_0)}{\lambda^{(2(1))}(t_0+2a,t_0)}\right)^{(t-t_s)/2a}\sim e^{-E_{D^*_0(D)}(t-t_s)/2a}$ compensates the residual time exponential dependence.
We do not take into account the contributions $\propto\, v_{\bar{\chi}_c \gamma_5 \chi_l}^{(2)}(t-t_s,t_0)$ and $v_{\bar{\chi}_c \chi_l}^{(1)}(t-t_s,t_0)$ to the projected correlators $\sum_i \langle \bar{\chi}_b \gamma_5
\chi_l(t_p) A_0(V_0) (t) O_i(\bar{\chi}_c, \chi_l,t_s)v^{(2(1))}_i(t-t_s,t_0)\rangle$
because the B meson goes through operator $A_0$ ($V_0$)
only to a pure scalar (pseudoscalar) state.

The ratio in eq.~\ref{ratioDscatoD} becomes
\beq\label{ratioDscatoDgevp}
\frac{\langle B|A_0|D^*_0 \rangle}{\langle B|V_0|D\rangle} &\simeq&
\frac{C^{(3)\prime}_{\rm B A_0  D^*_0}(t_p,t,t_s)}{C^{(3)\prime}_{\rm B V_0  D}(t_p,t,t_s)}\times
\frac{Z_A}{Z_V} 
\eeq
using eq~(\ref{eq:lambda})

\begin{figure*}[htbp]
\begin{center}
\includegraphics*[width=10cm,height=7cm] {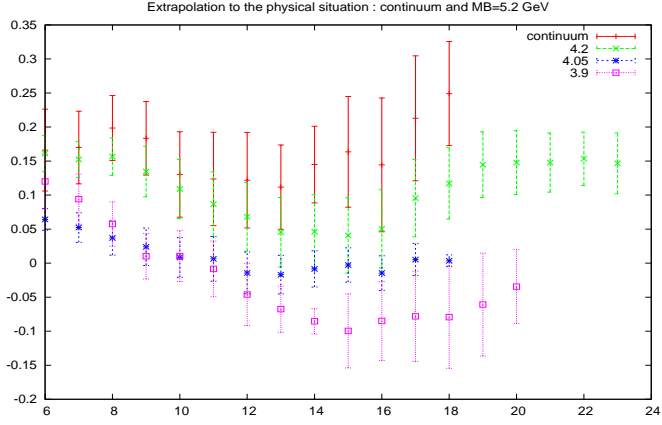}
\end{center}
\caption{\em The ratio in eq.~(\ref{ratioDscatoDgevp}) once symmetrised according 
to \eq{R5toC3} extrapoleted from the three $b$ quark masses to the physical
B mass : 5.2 GeV, linearly in $1/M_B$. We
have used data from $\beta = 3.9, 4.05, 4.2$. The time unit is the lattice spacing
for $\beta = 4.2$ : $a \simeq  0.054 \; {\rm fm}$.
The ratios for $\beta = 3.9, 4.05$ have been interpolated to points in 
$a_{\beta = 4.2}$ lattice units 
using the formula which is detailed in the caption of table 
\ref{tab:ratioDscalD}. The plot starts at $t = 6$, the time $t_0$ 
of the GEVP procedure. We see acceptable plateaus up to $t = 10$ and we
compute the averages between 6 and 10 which are reported in 
table~\ref{tab:ratioDscalD}. For larger times the signal falls down, presumably
because of the parity violation when using twisted masses which induces the
scalar $D^\star_0$ to ``decay'' into the pseudoscalar $D$ meson. 
The error bars are only statistical errors.} 
\label{fig:ratiossca}
\end{figure*}

Of course the ratio of branching fractions has to take into account the difference in phase space.
However, we ignore totally the dependence of the amplitude on the recoil, having only estimated the zero recoil contribution. Therefore, we will for the moment neglect the phase space dependence. 
We collect the results of eq.~\ref{ratioDscatoDgevp} in Table \ref{tab:ratioDscalD} and show plateaus
in Figure \ref{fig:ratiossca}. 

The values of the plateaus are reported in table~\ref{tab:ratioDscalD}.
The dependence in $m_B$ agrees with the formula $c/m_B+b$. We show both the 
extrapolation to the physical B meson and also to the vanishing lattice spacing.
When both extrapolations are combined we get a ratio of 0.17(6)(6).
This is a non vanishing signal, thanks to the data at $\beta=4.2$ which, lying closer
to the continuum, constrain more efficiently the continuum limit. 
 
As a gross estimate the ratio of branching fractions is the square 
of the ratio of amplitudes reported in Table \ref{tab:ratioDscalD}. 
The experimental value of the ratios of branching fractions 
can be very grossly estimated to be around 0.1-0.2 which would correspond to
a ratio of amplitudes $\sim 0.3-0.45$. Our estimate lies below this value,
but at this stage we must remain very careful : the experimental status 
of the $D^*_0$ is unclear, the resonance being very broad, and our theoretical 
estimate is affected by very large uncertainties.
Also the experimental situation is far from clear 
for the moment~\cite{Yaouanc:2014ypa}.

\begin{table}[htb]
\begin{center}
\begin{tabular}{||c|c|c|c|c||}
\hhline{|t:=====:t|}
$\boldsymbol{\beta}$ & \bfseries{ratio $\boldsymbol{m_{b(1)}}$} & \bfseries{ratio $\boldsymbol{m_{b(2)}}$} & \bfseries{ratio $\boldsymbol{m_{b(3)}}$} & ratio at physical B\\
\hhline{|:=:=:=:=:=:|}
3.9 &  O.23(3) & 0.17(2)& 0.11(3)& 0.06(4)  \\
\hhline{||-|-|-|-|-||}
4.05 & 0.20(2) &0.14(2) &0.08(4)& 0.03(5)\\ 
\hhline{||-|-|-|-|-||}
4.2 & 0.23(2) & 0.19(3) & 0.17(3)& 0.14(4) \\ 
\hhline{||-|-|-|-|-||}
continuum &0.22(5) & 0.17(5)& 0.17(7) &0.17(6)(6)\\ 
\hhline{|b:=====:b|}
\end{tabular}
\end{center}
\caption{\em We give the ratios defined in eq.~(\ref{ratioDscatoDgevp})
over a plateau corresponding to [6-10] in units of lattice spacing
for $\beta=4.2$. The lattice points at the other $\beta$'s do not match 
with those at $\beta=4.2$. To explain our method, let us take a point
$x_{4.2}$, we take for $\beta=3.9$ the two points $y_{3.9}$ and 
$y_{3.9}+1$ such that $y_{3.9}*a_{3.9}/a_{4.2}<x_{4.2} < 
(y_{3.9}+1)a_{3.9}/a_{4.2}$. Then from any function $f$ computed for 3.9
we define the interpolated function $f_{\rm res} = f(y_{3.9})((y_{3.9}+1)
a_{3.9}/a_{4.2}-x_{4.2})+f(y_{3.9}+1)(x_{4.2}-y_{3.9}a_{3.9}/a_{4.2})$.
The $b$ quarks range from the lightest to heaviest 
from left to right : $m_B \simeq 2.55,3.18,3.97$ GeV. The right column 
corresponds to the extrapolation at the physical $B$ mass : 5.2 GeV.
The last line corresponds to the extrapolation to the continuum. The 
errors are only statistical except on the physical case (bottom-right)
where the second error takes into account a systematic error estimated from 
the dependence of our result on the window choices to compute the
 masses and on the method in extrapolating to the physical point
 (in $1/m_B$ or in $1/\mu^{\overline {MS}}_b(2 GeV)$). }  
\label{tab:ratioDscalD}
\end{table}

\section{$\boldsymbol{B}$ decay to the tensor ($\boldsymbol{J=2^+}$) charmed meson}
\label{sec:BtoD2}
In this section we want to estimate the amplitudes for 
$B \to D^*_2 \ell \nu $ decay. 
In~\cite{Agashe:2014kda} the spin-two charmed mesons are 
 named  $D^*_2(2460)$ ($D^*_{s2}(2573)$).
For the initial $B$ meson, we use three ``$B_i,\ i=1,\,2,\,3$''
with increasing masses in the range 2.55, 3.18 and 3.97 GeV.
As was mentioned before, the ``$B_i,\ i=1,\,2,\,3$'' are moving while the final charmed meson 
is at rest. 
We concentrate on the calculation of the form factor
$\tilde k$ since it was shown in section~\ref{sec:estimation} that it is, by large, dominant 
in the decay width. 

\subsection{3-pt correlators computed for $\boldsymbol{B \to D(2^+)}$}
We start from the formulae recalled in section \ref{sec:estimation}. We use a symbolic notation to represent the hadronic matrix elements
\[
H_{i,j,k} = \langle B|A_k|D^{*}_{2}(\epsilon^{(\lambda)}_{ij})\rangle 
\qquad\leadsto\qquad\tpc{k}{j}{i}\,.
\]
The various combinations to extract $\tilde{k}$ are collected in Table \ref{tab:combiktilde}. We consider
all of these combinations and average the resulting value for $\tilde k$. To eliminate artefacts we must also apply the symmetrized result according to~\eq{R5toC3}

\begin{table}[htbp]
\begin{center}
\begin{tabular}{||c|c||}
\hhline{|t:==:t|}
{\bfseries combination} & {\bfseries expression} \\
\hhline{|:=:=:|}
\rule{0pt}{14pt}$ {p\,\tilde k} = -\,\sqrt{6}  \ampli{A}{1}{0}$ &  $\tpc{1}{1}{1} +  \tpc{1}{2}{2} - 2\,\tpc{1}{3}{3}$\\[2mm]	
${p\,\tilde k} = -\,\sqrt{6}\,\ampli{A}{2}{0} $ & $\tpc{2}{1}{1} + \tpc{2}{2}{2} - 2\,\tpc{2}{3}{3}$ \\[2mm]
${p\,\tilde k} = \,\sqrt{6}/2\,\ampli{A}{3}{0}$&   $ -\,(\tpc{3}{1}{1} + \tpc{3}{2}{2} - 2\,\tpc{3}{3}{3})/2$ \\[1mm] 
\hhline{|:=:=:|}
\rule{0pt}{14pt}${p\,\tilde k} = \left[{\ampli{A}{1}{+2} + \ampli{A}{1}{-2}}\right]$ &  $(\tpc{1}{1}{1} - \tpc{1}{2}{2})/2$ \\[2mm]
 ${p\,\tilde k} = -\,\left[{\ampli{A}{2}{+2} + \ampli{A}{2}{-2}}\right] $       &  $-\,(\tpc{2}{1}{1} - \tpc{2}{2}{2})/2$ \\[2mm]
\hhline{|:=:=:|}
\rule{0pt}{14pt}${p\,\tilde k} = {i}\,\left\{{\left[{\ampli{A}{1}{+2} - \ampli{A}{1}{-2}}\right] + \left[{\ampli{A}{1}{+1} + \ampli{A}{1}{-1}}\right]}\right\}$ & $\tpc{1}{1}{2}+\tpc{1}{2}{1} - \tpc{1}{3}{2}-\tpc{1}{2}{3}$\\[2mm]
${p\,\tilde k} =-{i}\,\left\{{\left[{\ampli{A}{3}{+2} - \ampli{A}{3}{-2}}\right] +\left[{\ampli{A}{3}{+1} + \ampli{A}{3}{-1}}\right]}\right\}$& $-\tpc{3}{1}{2}-\tpc{3}{2}{1}+\tpc{3}{3}{2}+\tpc{3}{2}{3}$\\[2mm]
\hhline{|:=:=:|}
\rule{0pt}{14pt}${p\,\tilde k} =i\,\left\{{\left[{\ampli{A}{1}{+1} + \ampli{A}{1}{-1}}\right] +\,i\, \left[{\ampli{A}{1}{+1} - \ampli{A}{1}{-1}}\right]}\right\}$ & $\tpc{1}{1}{3}+\tpc{1}{3}{1}- \tpc{1}{3}{2}-\tpc{1}{2}{3}$\\[2mm]
${p\,\tilde k}= -i\,\left\{{\left[{\ampli{A}{2}{+1} + \ampli{A}{2}{-1}}\right] +\,i\, \left[{\ampli{A}{2}{+1} - \ampli{A}{2}{-1}}\right]}\right\}$ & $-\tpc{2}{1}{3}-\tpc{2}{3}{1}+\tpc{2}{3}{2}+\tpc{2}{2}{3}$\\[2mm]
\hhline{|:=:=:|}
\rule{0pt}{14pt}${p\,\tilde k} = i\,\left\{{\left[{\ampli{A}{2}{+2} - \ampli{A}{2}{-2}}\right] +\,i\, \left[{\ampli{A}{2}{+1} - \ampli{A}{2}{-1}}\right]}\right\}$ & $\tpc{2}{1}{2}+\tpc{2}{2}{1}-\tpc{2}{3}{1}-\tpc{2}{1}{3}$\\[2mm]
${p\,\tilde k} = -i\,\left\{{\left[{\ampli{A}{3}{+2} - \ampli{A}{3}{-2}}\right] -\,i\, \left[{\ampli{A}{3}{+1} - \ampli{A}{3}{-1}}\right]}\right\}$ & $-\tpc{3}{1}{2}-\tpc{3}{2}{1}+\tpc{3}{3}{1}+\tpc{3}{1}{3}$\\[2mm]
\hhline{|b:==:b|}
\end{tabular}
\end{center}
\caption{\em Combinations of 3-pt correlators used to extract $\tilde{k}$.}
\label{tab:combiktilde}
\end{table}

\subsection {Subtracting zero momentum 3 point correlators}

The 3-pt correlation functions involving a tensor $D$ meson are unfortunately 
very noisy : hence it is extremely difficult to get 
a large enough signal-to-noise ratio. We will use a trick~\footnote{We 
are indebted to Philippe Boucaud who suggested this trick.} which consists
in subtracting to every 3-pt correlator the correlator with
 the same gauge configuration and the same operators at zero momentum. 
Indeed we know that the decay $B \to D^*_2$ vanishes at zero recoil. 
This is obvious in the continuum limit since we start with a $B$ meson 
of vanishing angular momentum $J$. The weak interaction operator 
(axial current $A_\mu$) having $J=0$ for $A_0$ and $J=1$ for $A_i\, (i=1,2,3)$,  
 cannot generate a $J=2$ state : at zero recoil there is no momentum to
 generate a higher angular momentum.\par 
However this vanishing is also exact on a lattice. The proof goes as follows:
the 3-pt correlators which contribute to the $ D^*_2 \to B$
 are linear combinations of correlators of the type
\beq\label{pirot}
C^{(3)}_{i,j,k}(t_p,t,t_s)\ =\ \langle O_{B}(t_p)
A_{k}(t)O_{D_i V_j}(t_s)\rangle\,.
\eeq    
where $i,j,k\in \{1,2,3\}$  may be different or equal. 
All operators are at rest (zero recoil). We have assumed the $D^*_2$ meson 
($B$ meson) interpolating field to be at the source time $t_s$ 
(sink time $t_p$), and the current at time $t$ with $t_p \ge t \ge t_s $.  \par
Let us choose one of the three spatial directions $\hat l$ and consider 
the rotation ${\mathscr R}_l(\pi)$ of angle $\pi$ around it : the spatial 
coordinates perpendicular to $\hat l$ change sign. All vector operators,
 $D_i, V_i, \text{and}\ A_i$ change sign if $i$ is perpendicular to $\hat l$ whereas they 
remain unchanged if $i=l$.\par
 ${\mathscr R}_l(\pi)$ belongs to the 3-D cubic symmetry group. The lattice 
actions are invariant under ${\mathscr R}_l(\pi)$, even the twisted mass action
since ${\mathscr R}_l(\pi)$ is parity even. In Eq.~\eqref{pirot}, there are three 
operators at three different times. Being at rest, we may assume that their 
spatial nesting is invariant under ${\mathscr R}_l(\pi)$ : it can be a stochastic
 source, a local operator at the origin of 3-space, a smeared operator
 symmetric around the origin of 3-space or a local operator integrated over
3-space (for the current). If an odd number among the indices $i,j,k$ are
 perpendicular to $\hat l$, then the correlator in \eq{pirot} changes sign 
under ${\mathscr R}_l(\pi)$ and the amplitude must vanish. This happens if
 $i=j=k$, $\hat l$ being any other direction or if $i= j \ne k$, $l=i$.\par
However, if $i,j,k$ are all different it does not work : any ${\mathscr R}_l(\pi)$
will keep the $C^{(3)}$ of \eq{pirot} unchanged and thus cannot be proven to vanish 
on the lattice although it should in the continuum limit. This type of term does 
generate lattice artefacts. A parity operation would change its sign (changing 
the sign of all three operators) but parity is not an invariant of the 
twisted mass action. We must then use correlators symmetrised according to 
the exact exact symetry of twisted mass action~\cite{Frezzotti:2003ni} i.e. 
apply \eq{R5toC3} :
the lattice artefact should then disappear and  
$C^{(3)\,\text{sym}}_{i,j,k}(t_s,t_c,t_p) = 0$ on the lattice, at zero recoil.\par
 Since it must vanish at zero recoil on the lattice, we may subtract to the 
three point correlator at non vanishing recoil the same configuration at zero
recoil. This reduces some correlated noise, and indeed it turns out that the
 signal, although still very noisy, is significantly improved.
We have computed the three point functions with both all $\mu$'s positive 
(set $sp_0$) and all opposite in sign (set $sp_1$). It turns out that the real
parts of the 3-point functions are very similar for both $sp_0$ and $sp_1$ 
sets, while the imaginary parts are approximatively opposite in signs, 
from which we can guess that the contributions with $i,j,k$ not all different
are dominantly real while the ones with  $i,j,k$ different are dominantly
imaginary. This is related to the fact that the terms odd in the $\mu$'s 
have an $i$ with respect to the ones which are even, but for the sake of
 brevity we will skip an exact proof.
\begin{figure*}[htbp]
\begin{center}
\includegraphics*[width=9cm, height=6.5cm]{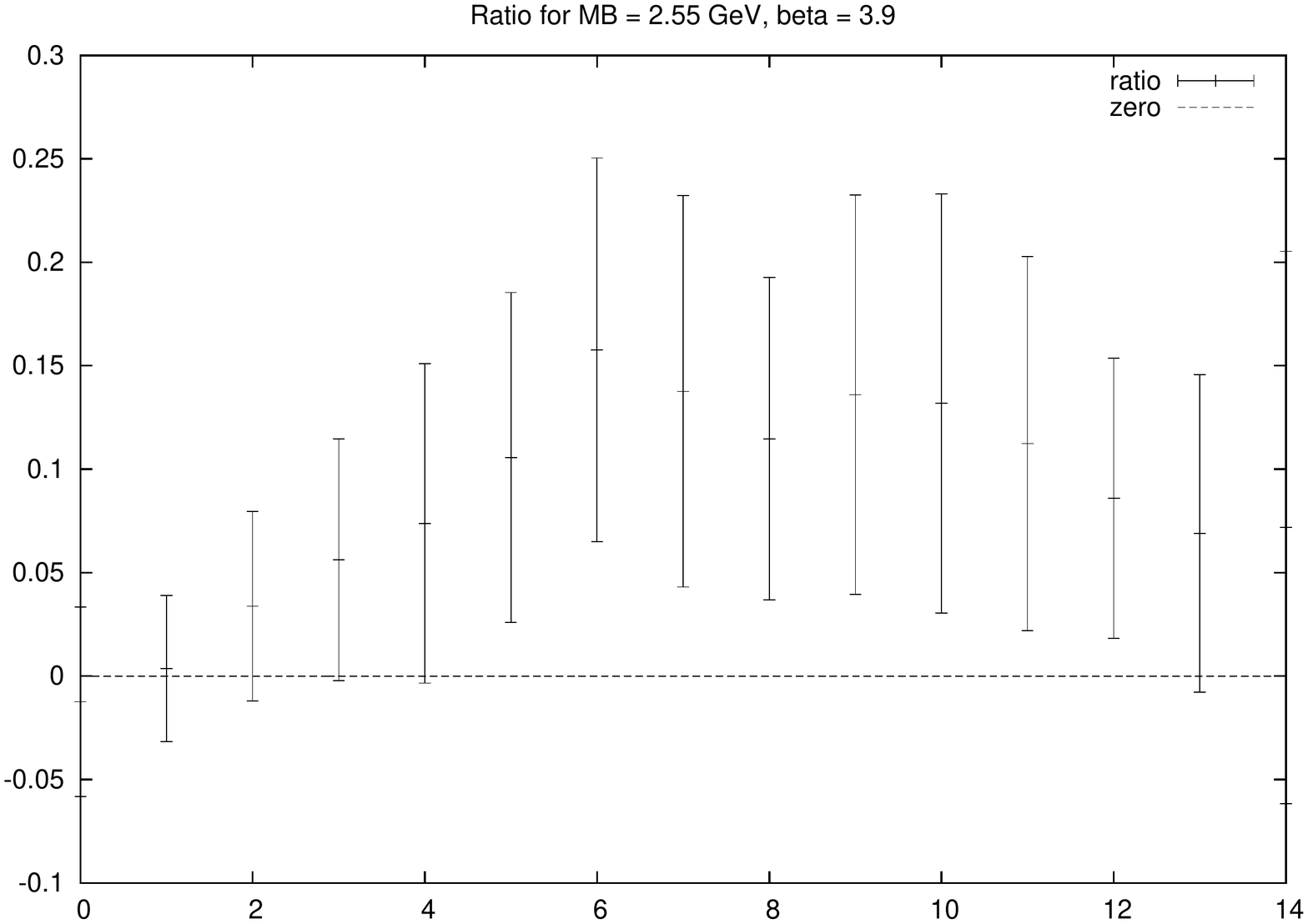}
\includegraphics*[width=9cm, height=6.5cm]{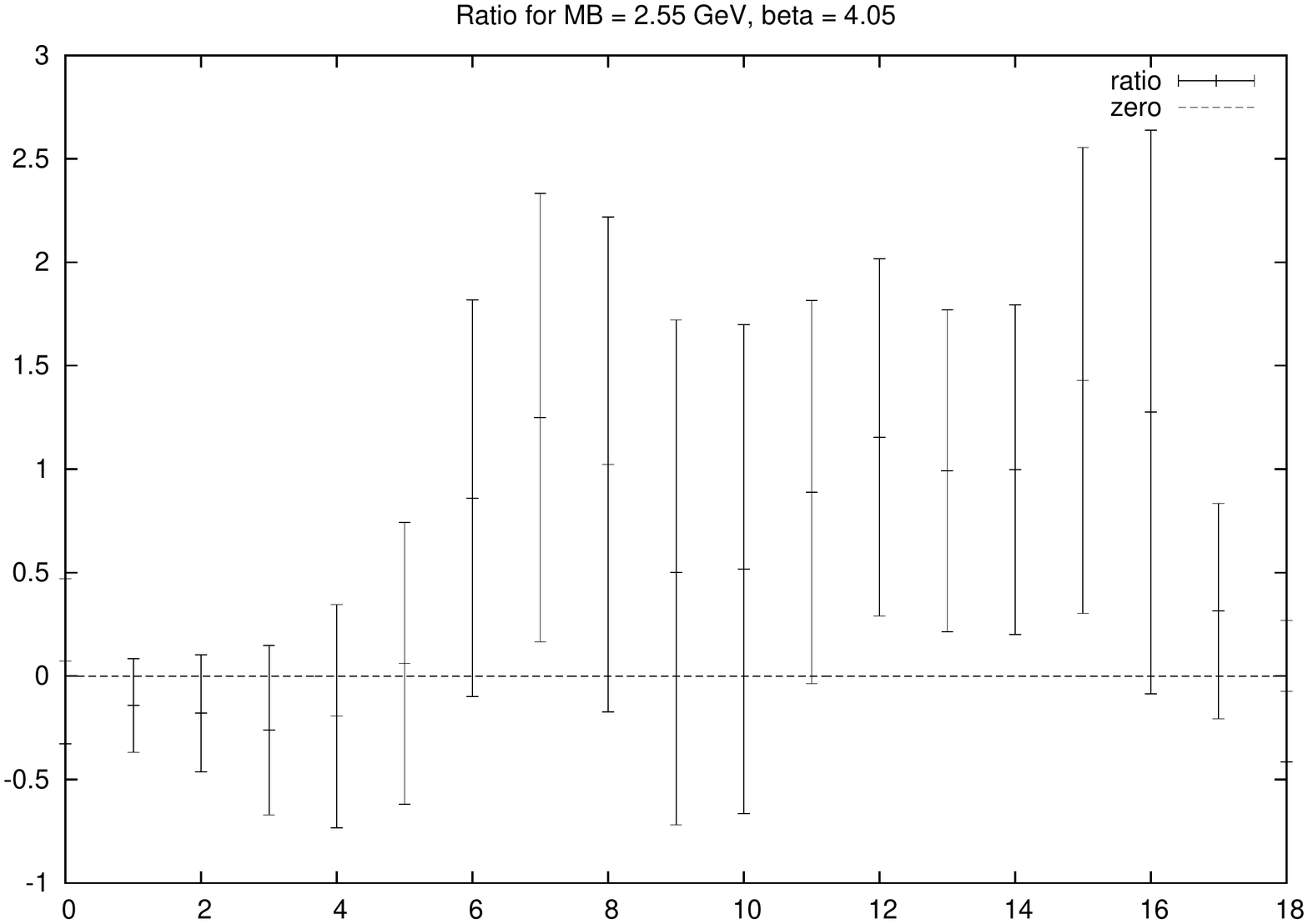}
\end{center}
\begin{center}
\includegraphics*[width=9cm, height=6.5cm]{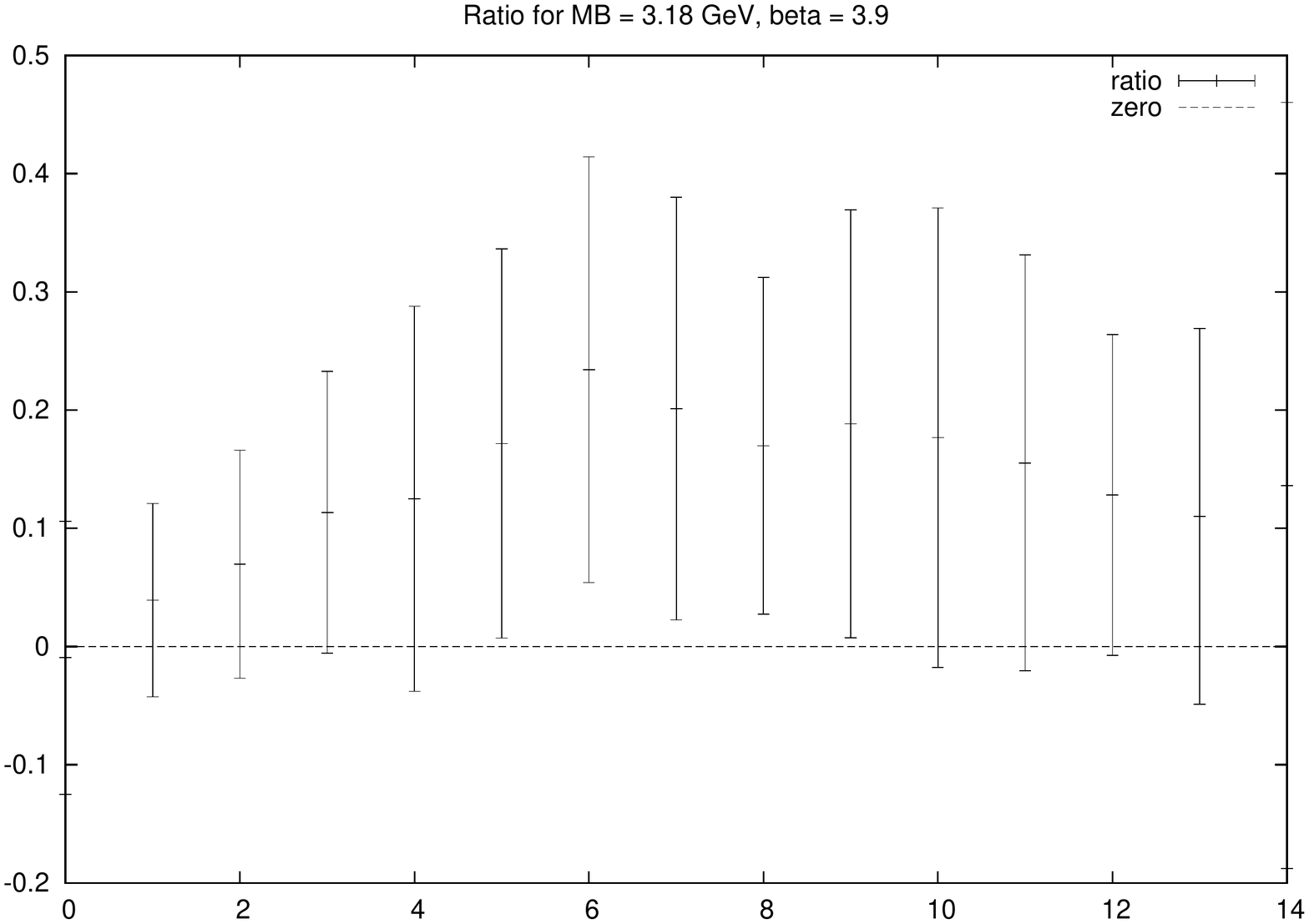}
\includegraphics*[width=9cm, height=6.5cm]{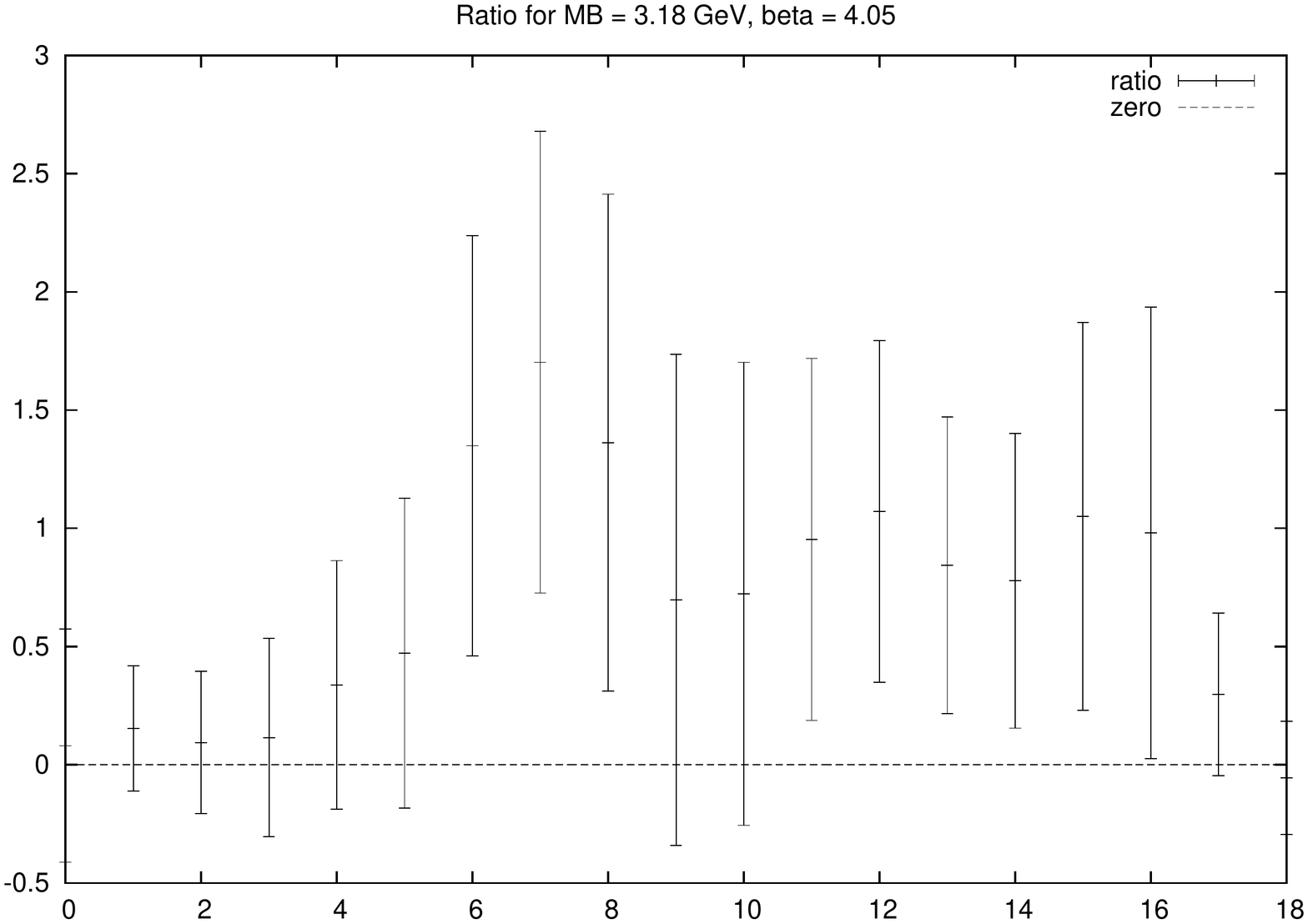}
\end{center}
\begin{center}
\includegraphics*[width=9cm, height=6.5cm]{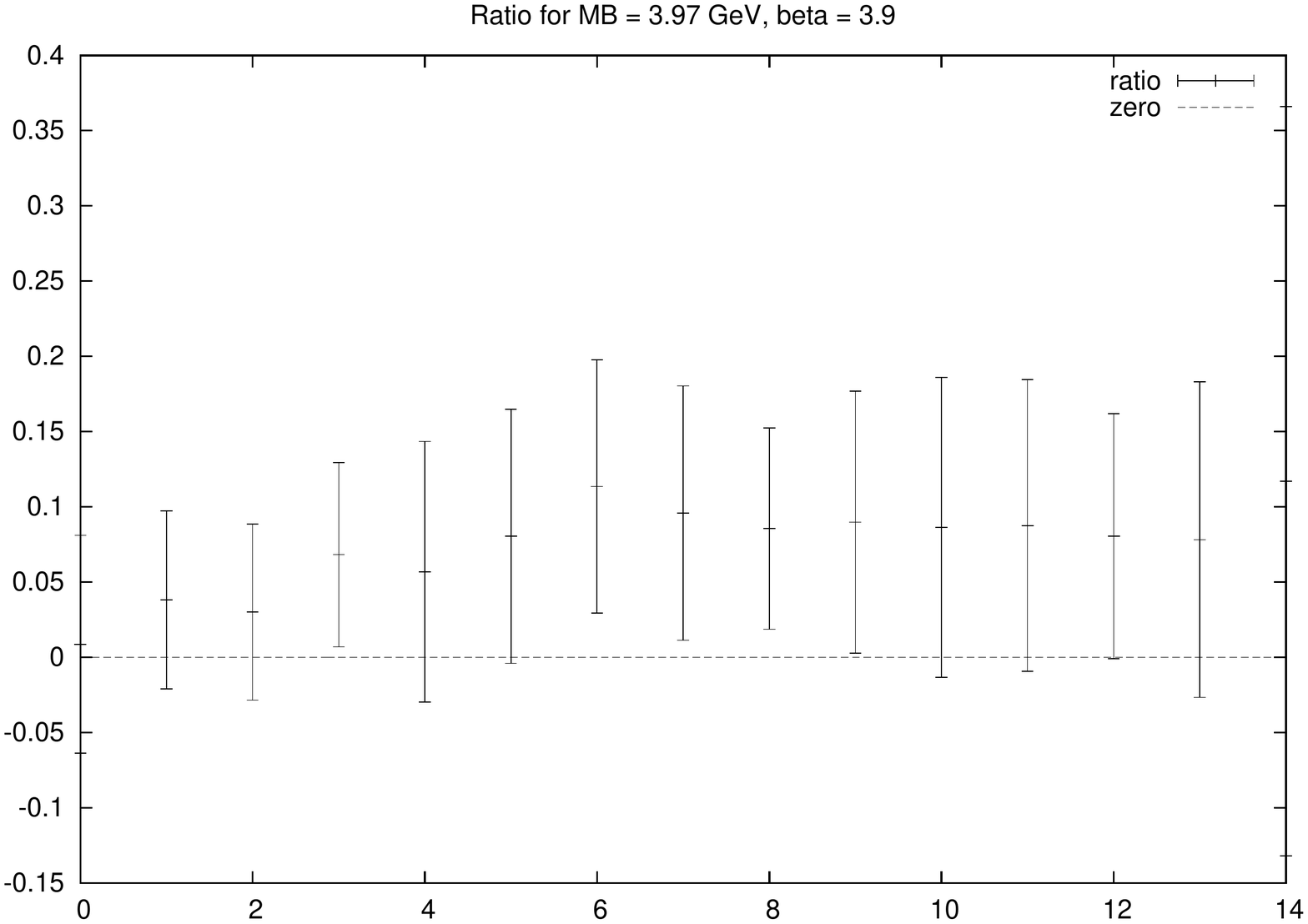}
\includegraphics*[width=9cm, height=6.5cm]{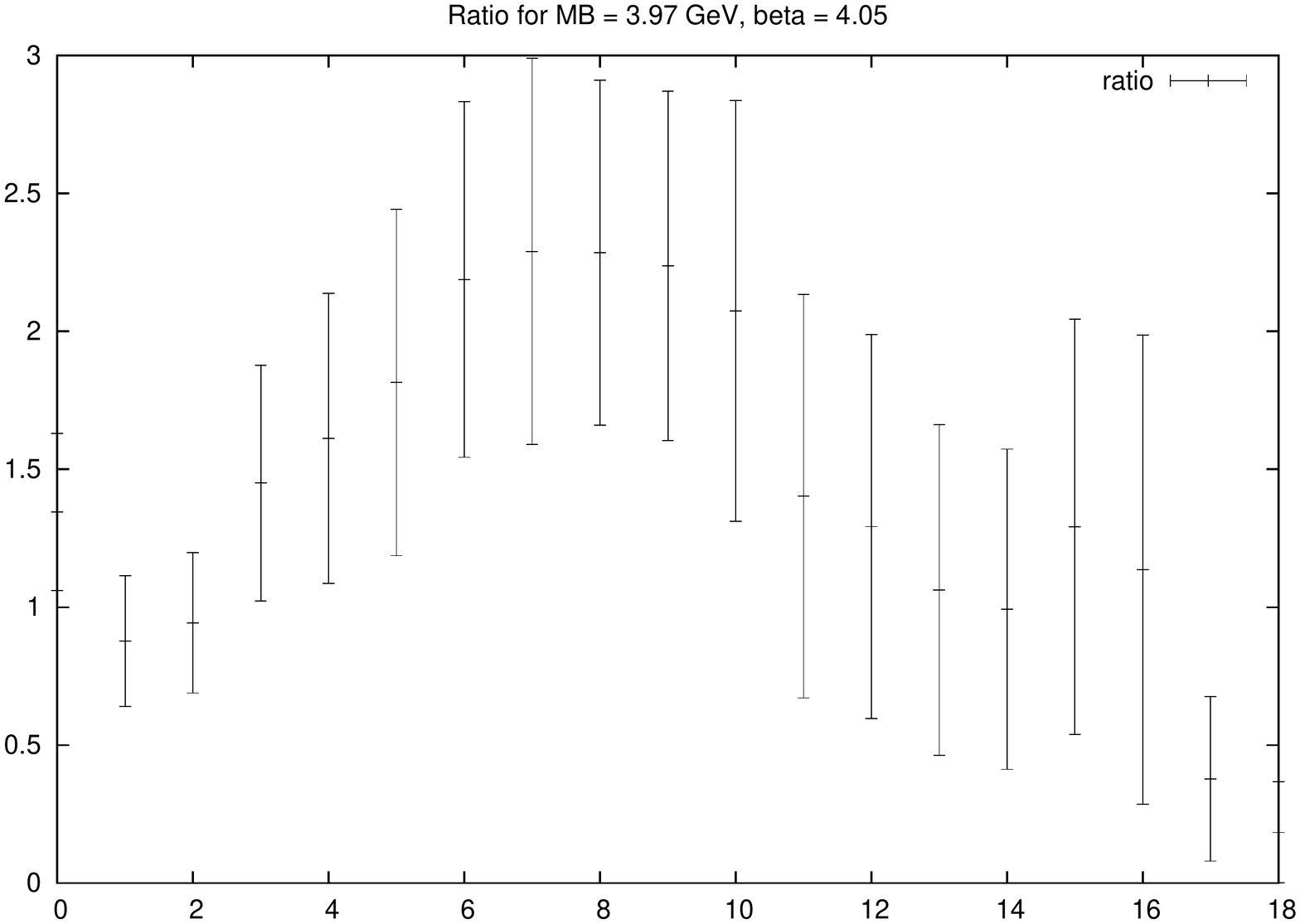}
\end{center}
\caption{\em The ratio of the matrix element for $B \to D^*_2$ over the
value derived from the infinite mass limit, \eq{eq:ratsinf}, for $w=1.3$,
 once the 3-pt function has 
been symmetrised according to \eq{R5toC3} and once the 3-pt function at
zero recoil has been subtracted. We show the three $b$ quark masses. 
The plots on the left correspond to  $\beta=3.9$, and $t_p-t_s = 14$,
 those to the right  to $\beta=4.05$ and $t_p-t_s = 18$.
The upper line corresponds to the B meson of continuum mass 2.55 GeV, the second
to 3.18 GeV and the third to 3.97 GeV. 
 We use for the three point functions the average of the 
combinations expanded in table~\ref{tab:combiktilde}.}
\label{fig:ratioB2D2}
\end{figure*}  
\subsection {Extracting the matrix element}

An estimate of $p \,\tilde k$ is thus given by:
\begin{equation}
R_{i,D^*_2,\vec{\theta}}(t_{p},t,t_s) = \frac{C^{(3)}_{B(\vec{\theta}) A_{i} D^*_2}(t_{p},t,t_s)
\sqrt{Z_{B}Z_{D^*_2}}\,Z_A}{C^{(2)}_{D^*_2}(t-t_s) C^{(2)}_{B(\vec{\theta})}(t_{p}-t)}\,,
\end{equation}
where the 3-pt correlators are the combinations of correlators listed in
table~\ref{tab:combiktilde}, symmetrised according to \eq{R5toC3} and with the
corresponding zero recoil 3-pt correlators subtracted.
Let us remind that 
$\vec p = \vec{\theta}\pi/L$ where $L$ is the spatial length of the lattice. We
have used systematically  $\vec{\theta}_x = \vec{\theta}_y =  \vec{\theta}_z =
\theta$ whence $|\vec p|= \sqrt{3}\, \theta\,\pi/L$. In fact we prefer to
present another ratio. Since the estimate of $B \to D^*_2$ in the infinite mass
limit is rather successful, in good agreement with experiment, we will compute
the ratio of the 3-pt correlators divided by the one which is derived from the
infinite mass limit formula~\cite{Morenas:1997nk}:
\begin{equation} 
\tau_{3/2}(w) = \tau_{3/2}(1)\left(\frac 2 {1+w} \right)^{2 \sigma^2_{3/2}}\,.
\end{equation}
Where the fit gives $\sigma^2_{3/2} \simeq 1.5$, and $ \tau_{3/2}(1) \simeq
0.54 $ and the formula in section 2.5.1
\beq\label{eq:kinf}
\tilde k_{\rm inf} = \sqrt 3 \, \sqrt {r_{D^*_2}} \,(1+w) \,\tau_{3/2}(w)\,,
\eeq 
with $r_{D^*_2} = M_{D^*_2}/M_B$.
This $\tilde k_{\rm inf}$ will be used as a benchmark for $\tilde k$ extracted 
from our present calculations. From our benchmark $\tilde k_{\rm inf}$ we
compute the benchmark three point correlator, the $D^*_2$ meson being created at
time $t_s$  the current inserted at time t and the $B$ anihilated at time $t_p$:
\beq\label{eq:infinite}
C^{(3)}_{\rm inf}(t_p,t,t_s) = \tilde k_{\rm inf} |\vec p|\,
 C^{(2)}_{D^*_2}(t-t_s)\,C^{(2)}_B(t_p-t)\,Z_A/\sqrt{Z_B Z_{D^*_2}}
\eeq 
We thus consider the ratio
\beq\label{eq:ratsinf}
\frac{\tilde k} {\tilde k_{\rm inf}} = \frac{C^{(3)}_{B(\vec{\theta})
 A_{i} D^*_2}(t_{p},t,t_s)}
{C^{(3)}_{\rm inf}(t_p,t,t_s)}
\eeq

To increase the signal we take the average on the 11 expressions for $\tilde k$
in Table~\ref{tab:combiktilde}, for all three  masses of the $B$ meson, 
for $\beta=3.9$ and $\beta=4.05$. For the 2-pt functions in \eq{eq:infinite} we 
have used the numerical values.
As can be seen all these plots show similar shapes. There is a positive noisy
signal, of the order 1 for $\beta=4.05$. For $\beta=3.9$ the ratio is about
one order of magnitude smaller. We do not understand this feature.

To conclude we may claim that there is a hint of a signal for $B \to D^\ast_2
 \ell \nu$ with finite $m_{c,b}$. But the size of the statistical and systmatic
 errors  do not allow us to provide any quantitative result.
 It is clear that improving the signal is a major goal. This can be performed
using larger statistics, using the points at $\beta=4.2$, using different times
for the sink, using other lattice actions, and trying to find better
interpolating fields. 

\section{Conclusions and prospects\label{sec:conclusion}}

The major goal of this paper concerned the orbitally excited states of the
charmed mesons and the semileptonic decay of the $B$ meson into the latter. 
We have concentrated on the $D^\ast_2$ and $D^\ast_0$ (see \cite{Wagner:2013laa}
for a study of the mass spectrum including the spin 1 particles).
We have considered three ``$B$ mesons" with respectively masses of 2.55, 3.18, 
3.97 GeV. We have used only two lattice spacings, 0.085 and 0.069 fm, with
 the addition of data with 0.054 fm for $D^\ast_0$.

Concerning the spectroscopy, we have noted a discrepancy between the masses of
the $D^\ast_2$ states which are in the $E$ ($D_i\,V_i$) discrete group and the 
ones in the $T_2$  ($D_i\,V_j; j \ne i$). This is of course a lattice artefact. 
We have also studied the $B$ meson energy as a function of the momentum.
 There is a clear departure from the theoretical formula for the heaviest meson.
 This is presumably also an artefact. 
 The $D^\ast_0$ state can decay into a $D$ meson due to the parity violation
 when using twisted quarks. It was necessary to use the GEVP method to overcome
 this difficulty. The mass differences between the $D^\ast_2$ ($D^\ast_0$) 
 with the $D$ meson mass, extrapolated to the continuum, do not agree well with
 experiment. There are indications that reducing the cut-off effects improves this result.
  
To compute the form factors and branching fractions we have derived all the
needed theoretical formulae necessary to estimate any form factor from lattice
calculation.  

Concerning $B \to D^\ast_0$ we have, up to now,
only considered the zero recoil quantity. Our result is that, contrary to the
case at infinite $b$ and $c$ masses, the zero recoil amplitude does not vanish.
This should increase drastically the ratio of $B \to D^\ast_0$ branching fraction 
over the $B \to D^\ast_2$ one, as compared to the infinite mass case. We estimate
the ratio of zero recoil amplitudes 
$(B \to Dast_0 l \nu)/(B \to D l \nu) = 0.17(6)(6)$. The corresponding ratio
of branching fractions should be around 0.02 with ery large errors.
Some experimental semileptonic branching ratios seem to indicate fort 
this ratio a figure of the order of 0.1, but the experimental situation 
is far from clear and our theoretical estimate needs still much work. 

The $B \to D^\ast_2$ is treated by a subtraction of the zero recoil contribution
which we prove to be theoretically vanishing. There is a signal, although
still very noisy. We take the infinite mass result as a benchmark. The ratio to
the infinite mass prediction is around 1 for $\beta = 4.05$ but around 0.1 
for $\beta = 3.9$ which indicates that beyond the very large statistical errors
the systematic effects are  not yet well understood.

Altogether, this paper has to be taken as a preliminary study. To our knowledge
it is the first study of semileptonic decays to orbitally excited charmed mesons
with finite masses for the $b$ and $c$. The considered process is very noisy 
and it is already rewarding that we got signals which seem to make sense,
although the uncertainty is still much too large. 

To improve the situation it seems that the path to follow is to further 
analyse the data set of ETMC at $\beta=4.2$ ($a \simeq 0.055$ fm)
and check against another lattice regularization. The
extrapolation to the continuum will thus be on a much safer ground. An increase
of the statistics might also help.

We also stress that $B_s$ and $D_s$ sectors are presumably interesting to examine. Indeed, the 
$D^*_{s0}(2317)$ and $D^*_{s1}(2460)$ states stand below the $DK$ and $D^* K$ thresholds: hence
they are narrow. At LHCb, according to a phenomenological study \cite{BecirevicTE}, about
100 events in the channel $B_s \to D^{*-}_{s0} \pi^+$ are expected with 1 ${\rm fb}^{-1}$ of integrated
luminosity. It is an encouragement to extend our effort of measuring the form factors of  
$B \to D^{**}$ semileptonic decays in the heavy-strange sector.

\section*{Acknowledgements} 
We acknowledge many enlightening discussions with Philippe Boucaud,
Francesco Sanfilippo and Marc Wagner. We have intensively used NISAN, the 
software created by F. Sanfilippo.
This work was granted access to the HPC resources of IDRIS under the 
allocation 2013-052271 made by GENCI. Many calculations were performed
on the computing center or IN2P3 in Lyon. 
Mariam Atoui acknwoledges the ``Conseil National de la Recherche 
Scientifique au Liban(CNRS), Liban" for her thesis grant.
\section*{Appendix}
\subsection*{Coefficients of the hadronic tensor $\boldsymbol{W_{\mu\nu}}$}
In order to compute explicitly the coefficients ${\alpha}$, ${\beta_{_{++}}}$, ${\beta_{_{+-}}}$, ${\beta_{_{-+}}}$, ${\beta_{_{--}}}$ and ${\gamma}$ given in~\eqref{eq:hadro}, we will have to evaluate the possible summation over the $D^{\ast\ast}$ spins.
\begin{description}
\item{\underline{\em Scalar meson} : } no summation and we get
\[
{
{
\begin{aligned}
&\boxed{\alpha} = 0\qquad\quad\boxed{\gamma} = 0\\
&\boxed{\beta_{_{++}}} = {\tilde u_+}^2\qquad\quad\boxed{\beta_{_{+-}}} = {\tilde u_+}{\tilde u_-}\qquad\quad\boxed{\beta_{_{-+}}} = {\tilde u_+}{\tilde u_-}\qquad\quad\boxed{\beta_{_{--}}} = {\tilde u_-}^2
\end{aligned}
}
}
\]
\item{\underline{\em Tensor meson $J=2$} : } the polarisation tensor $\poldb{\mu}{\nu}{p}$ satisfies~\cite{Spehler:1991yw,Choi:1992ba} :
\begin{multline*}
\sum_s \poldb{\mu}{\nu}{p}{}^\ast\,\poldb{\rho}{\sigma}{p} = -\,\dfrac13\left({\metrb{\mu}{\nu} - \dfrac{p_\mu\,p_\nu}{p^2}}\right)\left({\metrb{\rho}{\sigma} - \dfrac{p_\rho\,p_\sigma}{p^2}}\right)\ \\+
\ \dfrac12\left({\metrb{\mu}{\rho} - \dfrac{p_\mu\,p_\rho}{p^2}}\right)\left({\metrb{\nu}{\sigma} - \dfrac{p_\nu\,p_\sigma}{p^2}}\right)\ +
\ \dfrac12\left({\metrb{\mu}{\sigma} - \dfrac{p_\mu\,p_\sigma}{p^2}}\right)\left({\metrb{\nu}{\rho} - \dfrac{p_\nu\,p_\rho}{p^2}}\right)
\end{multline*}
After calculation, we obtain:
\end{description}
\[
{
{
\begin{aligned}
&\boxed{\alpha} = -\,\dfrac{(\pB\cdot\pDd)^2 - \mB^2\mDd^2}{2\,\mDd^2}\left[{
{\tilde k}^2 + 4\,{\tilde h}^2\Bigl({(\pB\cdot\pDd)^2 - \mB^2\mDd^2}\Bigr)
}\right]\\[4mm]
&\boxed{\beta_{_{++}}} = \begin{multlined}[t]
\dfrac{2\,\tilde k\,\tilde b_+}{3\,\mDd^4}\Bigl({\pB\cdot\pDd - \mDd^2}\Bigr)\Bigl({(\pB\cdot\pDd)^2 - \mB^2\mDd^2}\Bigr)
+ \dfrac{2\,{\tilde b_+}^2}{3\,\mDd^4}\Bigl({(\pB\cdot\pDd)^2 - \mB^2\mDd^2}\Bigr)^2\\
- \dfrac{{\tilde h}^2}{2\,\mDd^2}\Bigl({(\pB\cdot\pDd)^2 - \mB^2\mDd^2}\Bigr)(\pB - \pDd)^2
+ \dfrac{{\tilde k}^2}{24\,\mDd^4}\left[{
\mDd^2(\pB-\pDd)^2 + 4\Bigl({(\pB\cdot\pDd)^2 - \mB^2\mDd^2}\Bigr)
}\right]
\end{multlined}\\[4mm]
&\boxed{\beta_{_{+-}}} = \boxed{\beta_{_{-+}}} = \begin{multlined}[t]
\dfrac{2\,\tilde b_+\,\tilde b_-}{3\,\mDd^4}\Bigl({(\pB\cdot\pDd)^2 - \mB^2\mDd^2}\Bigr)^2
+ \dfrac{\tilde k\,\tilde b_-}{3\,\mDd^4}\Bigl({\pB\cdot\pDd - \mDd^2}\Bigr)\Bigl({(\pB\cdot\pDd)^2 - \mB^2\mDd^2}\Bigr)\\
- \dfrac{\tilde k\,\tilde b_+}{3\,\mDd^4}\Bigl({\pB\cdot\pDd + \mDd^2}\Bigr)\Bigl({(\pB\cdot\pDd)^2 - \mB^2\mDd^2}\Bigr)
+ \dfrac{{\tilde h}^2}{2\,\mDd^2}(\mB^2-\mDd^2)\Bigl({(\pB\cdot\pDd)^2 - \mB^2\mDd^2}\Bigr)\\
+ \dfrac{{\tilde k}^2}{24\,\mDd^4}\Bigl({3\,\mB^2\,\mDd^2 - 4\,(\pB\cdot\pDd)^2 + \mDd^4}\Bigr)
\end{multlined}
\\[4mm]
&\boxed{\beta_{_{--}}} = \begin{multlined}[t]
- \dfrac{2\,\tilde k\,\tilde b_-}{3\,\mDd^4}\Bigl({\pB\cdot\pDd + \mDd^2}\Bigr)\Bigl({(\pB\cdot\pDd)^2 - \mB^2\mDd^2}\Bigr)
+ \dfrac{2\,{\tilde b_-}^2}{3\,\mDd^4}\Bigl({(\pB\cdot\pDd)^2 - \mB^2\mDd^2}\Bigr)^2\\
- \dfrac{{\tilde h}^2}{2\,\mDd^2}\Bigl({(\pB\cdot\pDd)^2 - \mB^2\mDd^2}\Bigr)(\pB + \pDd)^2
+ \dfrac{{\tilde k}^2}{24\,\mDd^4}\left[{
\mDd^2(\pB+\pDd)^2 + 4\Bigl({(\pB\cdot\pDd)^2 - \mB^2\mDd^2}\Bigr)
}\right]
\end{multlined}\\[4mm]
&\boxed{\gamma} = \dfrac{\tilde k\,\tilde h}{\mDd^2}\Bigl({(\pB\cdot\pDd)^2 - \mB^2\mDd^2}\Bigr)
\end{aligned}
}
}
\]
\subsection*{Variation domains of $\boldsymbol{x}$ and $\boldsymbol{y}$}
{\bfseries{First type of constraints : $\boldsymbol{x = x(y)}$}}
\begin{description}
\item{\underline{\em Non-zero mass lepton} : }
\end{description}
\[
(\mL\neq 0)\qquad
\left\{
{
\begin{aligned}
&x_\text{max} = \dfrac{1}{2}\left\{{
1 + y - \rD^2 + \rL^2\left[{ 1 + \dfrac{1}{y}\left({1 - \rD^2}\right)}\right]
+ \left({1 - \dfrac{\rL^2}{y}}\right)\sqrt{\bigl[{y - (1-\rD)^2}\bigr]\bigl[{y-(1+\rD)^2}\bigr]}
}\right\}
\\[1mm]
&x_\text{min} = \dfrac{1}{2}\left\{{
1 + y - \rD^2 + \rL^2\left[{ 1 + \dfrac{1}{y}\left({1 - \rD^2}\right)}\right]
- \left({1 - \dfrac{\rL^2}{y}}\right)\sqrt{\bigl[{y - (1-\rD)^2}\bigr]\bigl[{y-(1+\rD)^2}\bigr]}
}\right\}
\end{aligned}
}\right.
\]
\[
\text{with :}\qquad\rL^2\leqslant y\leqslant \left(1 - \rD\right)^2
\]
\begin{description}
\item{\underline{\em Zero mass lepton} : }
\end{description}
\[
(\mL = 0)\qquad
\left\{
{
\begin{aligned}
&x_\text{max} = \dfrac{1}{2}\left[{
1 + y - \rD^2 + \sqrt{\bigl[{y - (1-\rD)^2}\bigr]\bigl[{y-(1+\rD)^2}\bigr]}
}\right]
\\[1mm]
&x_\text{min} = \dfrac{1}{2}\left[{
1 + y - \rD^2 - \sqrt{\bigl[{y - (1-\rD)^2}\bigr]\bigl[{y-(1+\rD)^2}\bigr]}
}\right]
\end{aligned}
}\right.
\]
\[
\text{with :}\qquad 0\leqslant y\leqslant \left(1 - \rD\right)^2
\]
{\bfseries{Second type of constraints : $\boldsymbol{y = y(x)}$}}
\begin{description}
\item{\underline{\em Non-zero mass lepton} : }
\end{description}
\[
(\mL\neq 0)\qquad
\left\{
{
\begin{aligned}
&y_\text{max} = \dfrac{1}{2}\left[{
x - \dfrac{\rD^2(x - 2\rL^2)}{1 - x + \rL^2} + \left({1 - \dfrac{\rD^2}{1-x+\rL^2}}\right)
\sqrt{x^2 - 4\rL^2}
}\right]
\\[1mm]
&y_\text{min} = \dfrac{1}{2}\left[{
x - \dfrac{\rD^2(x - 2\rL^2)}{1 - x + \rL^2} - \left({1 - \dfrac{\rD^2}{1-x+\rL^2}}\right)
\sqrt{x^2 - 4\rL^2}
}\right]
\end{aligned}
}\right.
\]
\[
\text{with :}\qquad2\,\rL\leqslant x\leqslant 1 - \rD^2 + \rL^2
\]
\begin{description}
\item{\underline{\em Zero mass lepton} : }
\end{description}
\[
(\mL = 0)\qquad
\left\{
{
\begin{aligned}
&y_\text{max} = x\left({1 - \dfrac{\rD^2}{1-x}}\right)
\\[1mm]
&y_\text{min} = 0
\end{aligned}
}\right.
\]
\[
\text{with :}\qquad 0\leqslant x\leqslant 1 - \rD^2 
\]
\subsection*{Expressions for the various decay widths}
{\bfseries{$\boldsymbol{\dfrac{\dd^2\Gamma}{\dd x\,\dd y}}$  differential decay width}}
\vspace*{5mm}\par\noindent
$\blacktriangleright\,$\fbox{\em $\etatp{3}{0}$ states} :
\begin{description}
\item{\underline{\em Non-zero mass lepton} : }
\[
\boxed{
\begin{aligned}
\dfrac{\dd^2\Gamma}{\dd x\,\dd y} = -\,\dfrac{G_F^2\,\abs{V_{cb}}^2}{128\pi^3}\,
\,\mB^5\,\Biggl\{{}\Biggr. &
- {\tilde u_+}^2\,\Bigl[{4\,\bigl[{x\,\rDz^2 + (1-x)(y-x)}\bigr]+\rL^2\bigl[{3\,y -4(x+\rDz^2)+\rL^2}\bigr]}\Bigr]\\[3mm]
&+ 2\,{\,{\tilde u_+}{\tilde u_-}\,}\,\rL^2\Bigl[{2(1-x-\rDz^2) + y + \rL^2}\Bigr]\\[3mm]
&+ {\tilde u_-}^2\,\rL^2\,(y - \rL^2)
\Biggl.{}\Biggr\}
\end{aligned}
}
\]
\item{\underline{\em Zero mass lepton} : }
\[
\boxed{
\dfrac{\dd^2\Gamma}{\dd x\,\dd y} = \dfrac{G_F^2\,\abs{V_{cb}}^2}{32\pi^3}\,
\,\mB^5\,
{\tilde u_+}^2\,\bigl[{x\,\rDz^2 + (1-x)(y-x)}\bigr]
}
\]
\end{description}
\par\noindent$\blacktriangleright\,$\fbox{\em $\etatp{3}{2}$ states} :
\begin{description}
\item{\underline{\em Non-zero mass lepton} : }
\[
\boxed{
\dfrac{\dd^2\Gamma}{\dd x\,\dd y} = -\,\dfrac{\mB}{256\pi^3}\,\dfrac{G_F^2\,\abs{V_{cb}}^2}{2}\,
\,\Biggl\{{}\Biggr.
C_1\,{\tilde k}^2 + C_2\,{\tilde h}^2 + C_3\,{\tilde b_+}^2 + C_4\,{\tilde b_-}^2 
+ 2\,C_5\,{\tilde k}\,{\tilde b_+}
+ 2\,C_6\,{\tilde k}\,{\tilde b}_-
+ 2\,C_7\,{\tilde b}_+\,{\tilde b}_-
+ 2\,C_8\,{\tilde h}\,{\tilde k}
\Biggl.{}\Biggr\}
}
\]
where the $C_i$ coefficients are given by :
\begin{align*}
&\boxed{C_1 = \begin{multlined}[t]
\dfrac{\mB^4}{3\,\rDd^4}\Biggl\{{}\Biggr.
\left[{y^2-(2+\rDd^2)\,y+(1-\rDd^2)^2}\right]\left[{2(1-x)(x-y) +\rDd^2(3\,y-2\,x)}\right] - 3\,y^2\,\rDd^4\\[1mm]
\hspace*{12mm}
- \rL^2\Bigl[{
(1-2\,x+y)\left[{2(1-y)^2-3\,\rDd^2(1+y)}\right] - \rDd^4(2\,x-\rDd^2)
}\Bigr]
\\[1mm]
\hspace*{90mm}
-2\,\rL^4\Bigl[{y^2-2(1+\rDd^2)\,y +1 - \rDd^2 + \rDd^4}\Bigr]
\Biggl.{}\Biggr\}
\end{multlined}
}\\[3mm]
&\boxed{C_2 = \begin{multlined}[t]
\dfrac{\mB^8}{\rDd^2}\left[y-(1-\rDd)^2\right]\left[y-(1+\rDd)^2\right]\\[1mm]
\hspace*{10mm}\times\Bigl[{}\Bigr.
\begin{multlined}[t]
y\left[{
2(1-x+\rDd)(1-x-\rDd) -(1-y+\rDd^2)(1+y-2\,x-\rDd^2)
}\right]\\[1mm]
\Bigl.{
+ \rL^2\left[{
(1+y-\rDd^2)(1+y-2\,x-\rDd^2) + 2\,\rL^2
}\right]
}\Bigr]
\end{multlined}
\end{multlined}
}\\[3mm]
&\boxed{C_3 = 
\dfrac{\mB^8}{6\,\rDd^4}\left[y-(1-\rDd)^2\right]^2\left[y-(1+\rDd)^2\right]^2
\Bigl[{
4\,x(1-x-\rDd^2)-4\,y(1-x) + \rL^2\left[{4(x+\rDd^2) - 3\,y -\rL^2}\right]
}\Bigr]
}\\[3mm]
&\boxed{C_4 = \dfrac{\mB^8}{6\,\rDd^4}\,\rL^2\,(y - \rL^2)\left[y-(1-\rDd)^2\right]^2\left[y-(1+\rDd)^2\right]^2}\\[3mm]
&\boxed{C_5 = \begin{multlined}[t]
\dfrac{\mB^6}{3\,\rDd^4}\left[y-(1-\rDd)^2\right]\left[y-(1+\rDd)^2\right]\\[1mm]
\times\Bigl[{
2(1-y-\rDd^2)\left[{(1-x)(x-y)-x\,\rDd^2}\right] -\rL^2\left[
{(1-y+\rDd^2)(1-3\,x+2\,y-\rDd^2+\rL^2) + 2\,x\,\rDd^2}
\right]
}\Bigr]
\end{multlined}
}\\[3mm]
&\boxed{C_6 = \begin{multlined}[t]
\dfrac{\mB^6}{3\,\rDd^4}\,\rL^2\,\left[y-(1-\rDd)^2\right]\left[y-(1+\rDd)^2\right]\\[1mm]
\times\left[{
(1-y+\rDd^2)(1-x+\rDd^2)+2\rDd^2(x-2) + \rL^2(1-y+\rDd^2)
}\right]
\end{multlined}
}\\[3mm]
&\boxed{C_7 = \dfrac{\mB^8}{6\,\rDd^4}\,\rL^2\,\left[y-(1-\rDd)^2\right]^2\left[y-(1+\rDd)^2\right]^2
\left[2(1 - x - \rDd^2) + y + \rL^2\right]}\\[3mm]
&\boxed{C_8 = -\,\dfrac{\mB^6}{\rDd^2}\left[y-(1-\rDd)^2\right]\left[y-(1+\rDd)^2\right]
\left[{y(1+y-2\,x-\rDd^2) + \rL^2(1+y-\rDd^2)}\right]}
\end{align*}
\item{\underline{\em Zero mass lepton} : }
We notice that the coefficients of $C_4$, $C_6$ and $C_7$ cancel in this limit, leading to :
\[
\boxed{
\dfrac{\dd^2\Gamma}{\dd x\,\dd y} = -\,\dfrac{\mB}{256\pi^3}\,\dfrac{G_F^2\,\abs{V_{cb}}^2}{2}\,
\,\Bigl[{}\Biggr.
C_1\,{\tilde k}^2 + C_2\,{\tilde h}^2 + C_3\,{\tilde b_+}^2
+ 2\,C_5\,{\tilde k}\,{\tilde b}_+ + 2\,C_8\,{\tilde h}\,{\tilde k}
\Biggl.{}\Bigr]
}
\]
where the $C_i$ coefficients are given by :
\begin{align*}
&\boxed{C_1 = 
\dfrac{\mB^4}{3\,\rDd^4}\Bigl[{}\Biggr.
\left[{y^2-(2+\rDd^2)\,y+(1-\rDd^2)^2}\right]\left[{2(1-x)(x-y) +\rDd^2(3\,y-2\,x)}\right] - 3\,y^2\,\rDd^4\Bigl.{}\Bigr]
}\\[3mm]
&\boxed{C_2 = \begin{multlined}[t]
\dfrac{\mB^8}{\rDd^2}\left[y-(1-\rDd)^2\right]\left[y-(1+\rDd)^2\right]\\
\hspace*{10mm}\times
y\left[{
2(1-x+\rDd)(1-x-\rDd) -(1-y+\rDd^2)(1+y-2\,x-\rDd^2)
}\right]
\end{multlined}
}\\[3mm]
&\boxed{C_3 = 
\dfrac{\mB^8}{6\,\rDd^4}\Bigl[y-(1-\rDd)^2\Bigr]^2\Bigl[y-(1+\rDd)^2\Bigr]^2
\Bigl[{
4\,x(1-x-\rDd^2)-4\,y(1-x)
}\Bigr]
}\\[3mm]
&\boxed{C_5 = 
\dfrac{\mB^6}{3\,\rDd^4}\Bigl[y-(1-\rDd)^2\Bigr]\Bigl[y-(1+\rDd)^2\Bigr]
\Bigl[{
2(1-y-\rDd^2)\left[{(1-x)(x-y)-x\,\rDd^2}\right]
}\Bigr]
}\\[3mm]
&\boxed{C_8 = -\,\dfrac{\mB^6}{\rDd^2}\Bigl[y-(1-\rDd)^2\Bigr]\Bigl[y-(1+\rDd)^2\Bigr]
{y\,(1+y-2\,x-\rDd^2) }}
\end{align*}
\end{description}
{\bfseries{$\boldsymbol{\dfrac{\dd\Gamma}{\dd y}}$ differential decay width}}\vspace*{2mm}\par\noindent
The form factors depend only on the $y$ parameter, but in an unknown way. So, the integration over the $x$ variable can be done through the use the expressions of the type $x = x(y)$.\bigskip\par\noindent
$\blacktriangleright\,$\fbox{\em $\etatp{3}{0}$ states} :
\begin{description}
\item{\underline{\em Non-zero mass lepton} : } 
\[
\boxed{
\dfrac{\dd\Gamma}{\dd y} = -\,\dfrac{G_F^2\,\abs{V_{cb}}^2}{128\pi^3}\,\mB^5\,
\,\Bigl[
D_1\,{\tilde u_+}^2 + 2\,D_2\,{\tilde u}_+\,{\tilde u}_- + D_3\,{\tilde u_-}^2
\Biggr]
}
\]
where the $D_i$ coefficients are function of $y$ and are given by :
\begin{align*}
&\boxed{D_1 = \begin{multlined}[t]
\dfrac{1}{3\,y^3}\,\left(y-\rL^2\right)^2\,\Bigl[{\left[y-(1-\rDz)^2\right]\left[y-(1+\rDz)^2\right]}\Bigr]^{1/2}\\[1mm]
\hspace*{20mm}
\times\Biggl\{{
2\,y\,{\left[y-(1-\rDz)^2\right]\left[y-(1+\rDz)^2\right]} + 
\rL^2\,\Bigl[{y^2-2\,y\,(1+\rDz^2)+4\,(1-\rDz^2)^2}\Bigr]
}\Biggr\}
\end{multlined}
}\\[3mm]
&\boxed{D_2 = 
\dfrac{1}{y^2}\,\rL^2\,\left(y-\rL^2\right)^2\,(1-\rDz^2)\,\Bigl[{\left[y-(1-\rDz)^2\right]\left[y-(1+\rDz)^2\right]}\Bigr]^{1/2}
}\\[3mm]
&\boxed{D_3 = 
\dfrac{1}{y}\,\rL^2\,\left(y-\rL^2\right)^2\,\Bigl[{\left[y-(1-\rDz)^2\right]\left[y-(1+\rDz)^2\right]}\Bigr]^{1/2}
}
\end{align*}
Let us recall that, in that expression of $\dd\Gamma/\dd y$, the $y$ parameter belongs in the interval : $\ \rL^2\leqslant y\leqslant (1-\rDz)^2$
\item{\underline{\em Zero mass lepton} : }
\[
\boxed{
\dfrac{\dd\Gamma}{\dd y} = -\,\dfrac{G_F^2\,\abs{V_{cb}}^2}{128\pi^3}\,\mB^5\,
\,{\tilde u_+}^2\,D_1 
}\qquad\text{where}\qquad\boxed{D_1 = \dfrac23\,\Bigl[{\left[y-(1-\rDz)^2\right]\left[y-(1+\rDz)^2\right]}\Bigr]^{3/2}}
\]
since the other coefficients $D_2$ and $D_3$ give zero in this case.
\end{description}
\eject
\par\noindent
$\blacktriangleright\,$\fbox{\em $\etatp{3}{2}$ states} :
\begin{description}
\item{\underline{\em Non-zero mass lepton} : } 
\[
\boxed{
\dfrac{\dd\Gamma}{\dd y} = -\,\dfrac{\mB}{256\pi^3}\,\dfrac{G_F^2\,\abs{V_{cb}}^2}{2}\,
\,\Biggl\{{}\Biggr.
D_1\,{\tilde k}^2 + D_2\,{\tilde h}^2 + D_3\,{\tilde b_+}^2 + D_4\,{\tilde b_-}^2 
+ 2\,D_5\,{\tilde k}\,{\tilde b}_+
+ 2\,D_6\,{\tilde k}\,{\tilde b}_-
+ 2\,D_7\,{\tilde b}_+\,{\tilde b}_-
+ 2\,D_8\,{\tilde h}\,{\tilde k}
\Biggl.{}\Biggr\}
}
\]
where the $D_i$ coefficients are given by :
\begin{align*}
&\boxed{D_1 = \begin{multlined}[t]
\dfrac{\mB^4}{\rDd^4}\,\dfrac{1}{9\,y^3}\,\left(y-\rL^2\right)^2\,\Bigl[{\left[y-(1-\rDd)^2\right]\left[y-(1+\rDd)^2\right]}\Bigr]^{3/2}\\[1mm]
\hspace*{20mm}
\times\Biggl\{{
y\,\Bigl[{y^2-2\,y\,(1-4\,\rDd^2)+(1-\rDd^2)^2}\Bigr] + 
\rL^2\,\Bigl[{2\,y^2-y\,(4-\rDd^2)+2\,(1-\rDd^2)^2}\Bigr]
}\Biggr\}
\end{multlined}
}\\[3mm]
&\boxed{D_2 = 
\dfrac{\mB^8}{\rDd^2}\,\dfrac{1}{3\,y^2}\,\left(y-\rL^2\right)^2\,(2\,y+\rL^2)\,\Bigl[{\left[y-(1-\rDd)^2\right]\left[y-(1+\rDd)^2\right]}\Bigr]^{5/2}
}\\[3mm]
&\boxed{D_3 = \begin{multlined}[t]
\dfrac{\mB^8}{\rDd^4}\,\dfrac{1}{18\,y^3}\,\left(y-\rL^2\right)^2\,\Bigl[{\left[y-(1-\rDd)^2\right]\left[y-(1+\rDd)^2\right]}\Bigr]^{5/2}\\[1mm]
\hspace*{20mm}
\times\Biggl\{{
2\,y\,{\Bigl[y-(1-\rDd)^2\Bigr]\Bigl[y-(1+\rDd)^2\Bigr]} + 
\rL^2\,\Bigl[{y^2-2\,y\,(1+\rDd^2)+4\,(1-\rDd^2)^2}\Bigr]
}\Biggr\}
\end{multlined}
}
\end{align*}
\begin{align*}
&\boxed{D_4 = 
\dfrac{\mB^8}{\rDd^4}\,\dfrac{1}{6\,y}\,\rL^2\,\left(y-\rL^2\right)^2\,\Bigl[{\left[y-(1-\rDd)^2\right]\left[y-(1+\rDd)^2\right]}\Bigr]^{5/2}
}\\[3mm]
&\boxed{D_5 = 
\dfrac{\mB^6}{\rDd^4}\,\dfrac{1}{18\,y^3}\,\left(y-\rL^2\right)^2\,\Bigl[{\left[y-(1-\rDd)^2\right]\left[y-(1+\rDd)^2\right]}\Bigr]^{5/2}
\Bigl[{
2\,y\,{\left(1 - y - \rDd^2\right)} + 
\rL^2\,\left({4 - y - 4\,\rDd^2}\right)
}\Bigr]
}\\[3mm]
&\boxed{D_6 = 
\dfrac{\mB^6}{\rDd^4}\,\dfrac{1}{6\,y^2}\,\rL^2\,\left(y-\rL^2\right)^2\,\Bigl[{\left[y-(1-\rDd)^2\right]\left[y-(1+\rDd)^2\right]}\Bigr]^{5/2}
}\\[3mm]
&\boxed{D_7 = 
\dfrac{\mB^8}{\rDd^4}\,\dfrac{1}{6\,y^2}\,\rL^2\,\left(y-\rL^2\right)^2\,(1-\rDd^2)\,\Bigl[{\left[y-(1-\rDd)^2\right]\left[y-(1+\rDd)^2\right]}\Bigr]^{5/2}
}\\[3mm]
&\boxed{D_8 = 0
}
\end{align*}
We recall that, in those formulae, the $y$ parameter varies inside the domain : $\ \rL^2\leqslant y\leqslant (1-\rDd)^2$\par\noindent
\item{\underline{\em Zero mass lepton} : }
\[
\boxed{
\dfrac{\dd\Gamma}{\dd y} = -\,\dfrac{\mB}{256\pi^3}\,\dfrac{G_F^2\,\abs{V_{cb}}^2}{2}\,
\,\Bigl[{}\Biggr.
D_1\,{\tilde k}^2 + D_2\,{\tilde h}^2 + D_3\,{\tilde b_+}^2
+ 2\,D_5\,{\tilde k}\,{\tilde b}_+ + 2\,D_8\,{\tilde h}\,{\tilde k}
\Biggl.{}\Bigr]
}
\]
where the $D_i$ coefficients are given by :
\begin{align*}
&\boxed{D_1 = 
\dfrac{\mB^4}{\rDd^4}\,\dfrac{1}{9}\,\Bigl[{\left[y-(1-\rDd)^2\right]\left[y-(1+\rDd)^2\right]}\Bigr]^{3/2}
\Bigl[{y^2-2\,y\,(1-4\,\rDd^2)+(1-\rDd^2)^2}\Bigr]
}\\[3mm]
&\boxed{D_2 = 
\dfrac{\mB^8}{\rDd^2}\,\dfrac{2}{3}\,y\,\Bigl[{\left[y-(1-\rDd)^2\right]\left[y-(1+\rDd)^2\right]}\Bigr]^{5/2}
}
\end{align*}
\begin{align*}
&\boxed{D_3 = 
\dfrac{\mB^8}{\rDd^4}\,\dfrac{1}{9}\,\Bigl[{\left[y-(1-\rDd)^2\right]\left[y-(1+\rDd)^2\right]}\Bigr]^{7/2}
}\hspace*{50mm}\\[3mm]
&\boxed{D_5 = 
\dfrac{\mB^6}{\rDd^4}\,\dfrac{1}{9}\,(1-y-\rDd^2)\,\Bigl[{\left[y-(1-\rDd)^2\right]\left[y-(1+\rDd)^2\right]}\Bigr]^{5/2}
}\\[3mm]
&\boxed{D_8 = 0
}
\end{align*}
Here, the $y$ parameter lies in the domain : $\ 0\leqslant y\leqslant (1-\rDd)^2$\par\noindent
\end{description}
{\bfseries{$\boldsymbol{\dfrac{\dd\Gamma}{\dd x}}$ differential decay width}}\vspace*{2mm}\par\noindent
It is impossible to give general expressions for the leptonic spectra $\dfrac{\dd\Gamma}{\dd x}$ since the integration over $y$ can not be performed because we do not know the dependance of the form factors on $y$.\par\noindent
Nevertheless, the procedure to do those calculations is the following :
\begin{enumerate}
\item{} We start from the expressions of the $\dfrac{\dd^2\Gamma}{\dd x\,\dd y}$ decay widths given above
\item{} We use the constraints of the type $y = y(x)$ in order to perform the integration over $y$ from $y_\text{min}$ to $y_\text{max}$ (expressions given also above). Incidently, we must not forget that the maximum of $\ED$ corresponds to the minimum of $y$ and vice-versa. So, to integrate over $\ED$ from $\ED^\text{ min}$ to $\ED^\text{ max}$, we have to integrate equivalently over $y$ from $y_\text{max}$ to $y_\text{min}$ :
\[
\dfrac{\dd\Gamma}{\dd x} = \int\limits_{\mathclap{y_\text{max}}}^{\mathclap{y_\text{min}}}\dfrac{\dd^2\Gamma}{\dd x\,\dd y}\,\dd y
\]
with :
\[
\left\{
{
\begin{aligned}
&y_\text{max} = \dfrac{1}{2}\left[{
x - \dfrac{\rD^2(x - 2\rL^2)}{1 - x + \rL^2} + \left({1 - \dfrac{\rD^2}{1-x+\rL^2}}\right)
\sqrt{x^2 - 4\rL^2}
}\right]
\\[1mm]
&y_\text{min} = \dfrac{1}{2}\left[{
x - \dfrac{\rD^2(x - 2\rL^2)}{1 - x + \rL^2} - \left({1 - \dfrac{\rD^2}{1-x+\rL^2}}\right)
\sqrt{x^2 - 4\rL^2}
}\right]
\end{aligned}
}\right.
\]
($\rL = 0$ gives the relations in the case of a zero mass lepton.)
\item{} The last free parameter $x$ lies in the domain :
\[
2\,\rL\leqslant x\leqslant 1 - \rD^2 + \rL^2
\]
(Once again, $\rL = 0$ gives the variation domain in the case of a zero mass lepton.)
\end{enumerate}
{\bfseries{Total decay width $\boldsymbol{\Gamma}$}}\vspace*{2mm}\par\noindent
The problem, mentioned for the leptonic spectra, pops up here again because, in order to get $\Gamma$, we will have to integrate over $y$ at some point. So we will have to follow the same procedure.
\subsection*{Polarization tensor for the $\boldsymbol{\etatp{3}{2}}$ state}
Using expressions for the spin-1 polarisation vector found in~\cite{wein} for instance and the values of the Clebsch-Gordan coefficient from the ``Particle Physics Booklet'', we get :
\begin{gather*}
\poldh{\mu}{\nu}{+2} = \dfrac12\,\begin{pmatrix*}
0&0&0&0\\0&1&i&0\\0&i&-1&0\\0&0&0&0
\end{pmatrix*}
\qquad
\poldh{\mu}{\nu}{-2} = \dfrac12\,\begin{pmatrix*}
0&0&0&0\\0&1&-i&0\\0&-i&-1&0\\0&0&0&0
\end{pmatrix*}
\qquad
\poldh{\mu}{\nu}{+1} = \dfrac12\,\begin{pmatrix*}
0&0&0&0\\0&0&0&-1\\0&0&0&-i\\0&-1&-i&0
\end{pmatrix*}
\\[2mm]
\poldh{\mu}{\nu}{-1} = \dfrac12\,\begin{pmatrix*}
0&0&0&0\\0&0&0&1\\0&0&0&-i\\0&1&-i&0
\end{pmatrix*}
\qquad
\poldh{\mu}{\nu}{0} = \dfrac{1}{\sqrt{6}}\,\begin{pmatrix*}
0&0&0&0\\0&-1&0&0\\0&0&-1&0\\0&0&0&2
\end{pmatrix*}
\end{gather*}
(We dropped the $\vec 0$ in the notation.)
\subsection*{Extraction of the form factors}
The following expressions are not exhaustive.\par\noindent
Note that it is possible to recover the momentum transfer $y\,\mB^2 = (\pB - \pD)^2$ using :
\[
\EB =  \dfrac{\mB}{2\,\rD}\,[1 - y + \rD^2] \qquad\text{and}\qquad p^2 = \dfrac{\mB^2}{12\,\rD^2}\Bigl[{y - (1 - \rD)^2}\Bigr]\,\Bigl[{y - (1 + \rD)^2}\Bigr]
\]
leading to :
\[
\EB + \mD = \dfrac{\mB}{2\,\rD}(1-y+3\,\rD^2)\qquad\qquad\quad\EB - \mD = \dfrac{\mB}{2\,\rD}(1-y-\rD^2)
\]
\fbox{\bfseries{$\boldsymbol{\etatp{3}{0}}$ form factors}}\vspace*{2mm}\par\noindent
With $\quad\amplib{A}{\mu}\overset{\text{def.}}{=} \brakket[\big]{\etatp{3}{0}}{A_\mu}{B(\pB)}$
\begin{itemize}
\item{\underline{form factor $\tilde u_+$} :}
\[
\boxed{\tilde u_+} = -\,\dfrac{1}{2\,\mD}\left[{\dfrac{\EB - \mD}{p}\amplib{A}{i} - \amplib{A}{0}}\right] = -\,\dfrac{1}{2\,\mD}\left[{\dfrac{\EB - \mD}{3\,p}(\amplib{A}{1}+\amplib{A}{2}+\amplib{A}{3}) - \amplib{A}{0}}\right]
\]
\item{\underline{form factor $\tilde u_-$} :}
\[
\boxed{\tilde u_-} = \dfrac{1}{2\,\mD}\left[{\dfrac{\EB + \mD}{p}\amplib{A}{i} - \amplib{A}{0}}\right] = \dfrac{1}{2\,\mD}\left[{\dfrac{\EB + \mD}{3\,p}(\amplib{A}{1}+\amplib{A}{2}+\amplib{A}{3}) - \amplib{A}{0}}\right]
\]
\end{itemize}\vspace*{2mm}\par\noindent
\fbox{\bfseries{$\boldsymbol{\etatp{3}{2}}$ form factors}}\vspace*{2mm}\par\noindent
With $\quad
\amplia{A}{\mu}{\lambda}\overset{\text{def.}}{=} \brakket[\big]{\etatp{3}{2}(\lambda)}{A_\mu}{B(\pB)}\qquad\text{and}\qquad
\amplia{V}{\mu}{\lambda}\overset{\text{def.}}{=} \brakket[\big]{\etatp{3}{2}(\lambda)}{V_\mu}{B(\pB)}
$
\begin{itemize}
\item{\underline{form factor $\tilde k$} :}
\[
{
\boxed{\tilde k} = -\,\dfrac{\sqrt{6}}{p}\,\amplia{A}{1}{0} = -\,\dfrac{\sqrt{6}}{p}\,\amplia{A}{2}{0} = \dfrac{\sqrt{6}}{2\,p}\,\amplia{A}{3}{0} = \dfrac{1}{p}\,\left[{\amplia{A}{1}{+2} + \amplia{A}{1}{-2}}\right] = -\,\dfrac{1}{p}\,\left[{\amplia{A}{2}{+2} + \amplia{A}{2}{-2}}\right]
}
\]
\item{\underline{form factors $\tilde b_+$ and $\tilde b_-$} :}
\begin{gather*}
\left\{
\begin{aligned}
\boxed{\tilde b_{+}} &= -\,\dfrac{1+i}{4}\,\dfrac{1}{p^3\,\mD}\Bigl[{(\EB - \mD)(i\,\amplia{A}{1}{+2} + \amplia{A}{2}{+2}) - p(1+i)\amplia{A}{0}{+2}}\Bigr]\\[2mm]
\boxed{\tilde b_{-}} &= \dfrac{1+i}{4}\,\dfrac{1}{p^3\,\mD}\Bigl[{(\EB + \mD)(i\,\amplia{A}{1}{+2} + \amplia{A}{2}{+2}) - p(1+i)\amplia{A}{0}{+2}}\Bigr]
\end{aligned}
\right.
\\[3mm]
\left\{
\begin{aligned}
\boxed{\tilde b_{+}} &= \dfrac{1-i}{4}\,\dfrac{1}{p^3\,\mD}\Bigl[{(\EB - \mD)(i\,\amplia{A}{1}{-2} - \amplia{A}{2}{-2}) + p(1-i)\amplia{A}{0}{-2}}\Bigr]\\[2mm]
\boxed{\tilde b_{-}} &= -\,\dfrac{1-i}{4}\,\dfrac{1}{p^3\,\mD}\Bigl[{(\EB + \mD)(i\,\amplia{A}{1}{-2} - \amplia{A}{2}{-2}) + p(1-i)\amplia{A}{0}{-2}}\Bigr]
\end{aligned}
\right.
\\[3mm]
\left\{
\begin{aligned}
\boxed{\tilde b_{+}} &= \dfrac{1+i}{4}\,\dfrac{1}{p^3\,\mD}\Bigl[{(\EB - \mD)(\amplia{A}{1}{+1} +\amplia{A}{2}{+1}-\amplia{A}{3}{+1}) - p\,\amplia{A}{0}{+1}}\Bigr]\\[2mm]
\boxed{\tilde b_{-}} &= -\,\dfrac{1+i}{4}\,\dfrac{1}{p^3\,\mD}\Bigl[{(\EB + \mD)(\amplia{A}{1}{+1} +\amplia{A}{2}{+1}-\amplia{A}{3}{+1}) - p\,\amplia{A}{0}{+1}}\Bigr]
\end{aligned}
\right.
\\[3mm]
\left\{
\begin{aligned}
\boxed{\tilde b_{+}} &= -\,\dfrac{1-i}{4}\,\dfrac{1}{p^3\,\mD}\Bigl[{(\EB - \mD)(\amplia{A}{1}{-1} +\amplia{A}{2}{-1}-\amplia{A}{3}{-1}) - p\,\amplia{A}{0}{-1}}\Bigr]\\[2mm]
\boxed{\tilde b_{-}} &= \dfrac{1-i}{4}\,\dfrac{1}{p^3\,\mD}\Bigl[{(\EB + \mD)(\amplia{A}{1}{-1} +\amplia{A}{2}{-1}-\amplia{A}{3}{-1}) - p\,\amplia{A}{0}{-1}}\Bigr]
\end{aligned}
\right.
\\[3mm]
\left\{
\begin{aligned}
\boxed{\tilde b_{+}} &= \dfrac{1}{2\,i}\,\dfrac{1}{p^3\,\mD}\Bigl[{(\EB - \mD)\amplia{A}{3}{+2} - p\,\amplia{A}{0}{+2}}\Bigr]\\[2mm]
\boxed{\tilde b_{-}} &= \dfrac{1}{2\,i}\,\dfrac{1}{p^3\,\mD}\Bigl[{-\,(\EB + \mD)\amplia{A}{3}{+2} + p\,\amplia{A}{0}{+2}}\Bigr]
\end{aligned}
\right.
\end{gather*}
\item{\underline{form factor $\tilde h$} :}
\[
\boxed{\tilde h}\ =\ \dfrac{1}{2\,i}\,\dfrac{1}{p\,\mD}\dfrac{\amplia{V}{1}{\lambda}}{\poldh{\ast 3}{\alpha}{\lambda}\,{\pB}_{\!\alpha}-\poldh{\ast 2}{\alpha}{\lambda}\,{\pB}_{\!\alpha}}
\ =\ \dfrac{1}{2\,i}\,\dfrac{1}{p\,\mD}\dfrac{\amplia{V}{2}{\lambda}}{\poldh{\ast 1}{\alpha}{\lambda}\,{\pB}_{\!\alpha}-\poldh{\ast 3}{\alpha}{\lambda}\,{\pB}_{\!\alpha}}
\ =\ -\dfrac{1}{2\,i}\,\dfrac{1}{p\,\mD}\dfrac{\amplia{V}{3}{\lambda}}{\poldh{\ast 2}{\alpha}{\lambda}\,{\pB}_{\!\alpha}-\poldh{\ast 1}{\alpha}{\lambda}\,{\pB}_{\!\alpha}}
\]
where
\begin{center}
\begin{tabular}{||c||c|c|c||}
\hhline{|t:=:t:===:t|}
\rule{0pt}{14pt}$\boldsymbol{\lambda}$ & $\boldsymbol{\poldh{\ast 3}{\alpha}{\lambda}\,{\pB}_{\!\alpha}-\poldh{\ast 2}{\alpha}{\lambda}\,{\pB}_{\!\alpha}}$
& $\boldsymbol{\poldh{\ast 1}{\alpha}{\lambda}\,{\pB}_{\!\alpha}-\poldh{\ast 3}{\alpha}{\lambda}\,{\pB}_{\!\alpha}}$
& $\boldsymbol{\poldh{\ast 2}{\alpha}{\lambda}\,{\pB}_{\!\alpha}-\poldh{\ast 1}{\alpha}{\lambda}\,{\pB}_{\!\alpha}}$\\[5pt]
\hhline{|:=::=:=:=:|}
\rule{0pt}{15pt}+2&$-\,\dfrac{p}{2}(1+i)$&$-\,\dfrac{p}{2}(1-i)$&$p$\\[5pt]
\hhline{||-||-|-|-||}
\rule{0pt}{15pt}+1&$\dfrac{p}{2}$&$i\,\dfrac{p}{2}$&$-\,\dfrac{p}{2}(1+i)$\\[5pt]
\hhline{||-||-|-|-||}
\rule{0pt}{17pt}0&$-\,p\,\sqrt{\dfrac{3}{2}}$&$p\,\sqrt{\dfrac{3}{2}}$&$0$\\[6pt]
\hhline{||-||-|-|-||}
\rule{0pt}{15pt}-1&$-\,\dfrac{p}{2}$&$i\,\dfrac{p}{2}$&$\dfrac{p}{2}(1-i)$\\[5pt]
\hhline{||-||-|-|-||}
\rule{0pt}{15pt}-2&$-\,\dfrac{p}{2}(1-i)$&$-\,\dfrac{p}{2}(1+i)$&$p$\\[5pt]
\hhline{|b:=:b:===:b|}
\end{tabular}
\end{center}
\end{itemize}
\subsection*{Masses and energies}
We collect in Table~\ref{tabmassenergies} the masses and energies that we 
extract in our analysis.\par\noindent
\begin{table}[htbp]
\begin{center}
\begin{tabular}{||c||c|c||c|c||c|c||}
\hhline{~|t:==:t:==:t:==:t|}
\multicolumn{1}{l||}{}&\multicolumn{2}{c||}{$\boldsymbol{\beta=3.9}$}&
\multicolumn{2}{c||}{$\boldsymbol{\beta=4.05}$}&
\multicolumn{2}{c||}{$\boldsymbol{\beta=4.2}$}\\
\hhline{|t:=:|--||--||--||}
\bfseries{meson}&$\boldsymbol{\theta}$&$\boldsymbol{E(\theta)}$&$\boldsymbol{\theta}$&$\boldsymbol{E(\theta)}$
&$\boldsymbol{\theta}$&$\boldsymbol{E(\theta)}$\\
\hhline{|:=::=:=::=:=::=:=:|}
$D$&0&0.76(1)&0&0.62(1)&0 & 0.52(1)\\
$D^*_0$&0&1.01(4)&0&0.80(2)&0 &0.66(1)\\
$D^*_2$&0&1.14(2)&0&0.93(2)& - & -\\
\hhline{||-||-|-||-|-||-|-||}
$B(\mu_{h_1})$&0&1.00(1)&0&0.82(1)& 0&0.69(1)\\
$B(\mu_{h_2})$&0&1.21(1)&0&1.01(1)& 0&0.85(1)\\
$B(\mu_{h_3})$&0&1.50(1)&0&1.25(1)& 0&1.05(1)\\
\hhline{|:=::==::==::==:|}
$B(\mu_{h_1})$&0.99&1.02(1)&1.09&0.84(1\\
$B(\mu_{h_2})$&1.21&1.24(1)&1.35&1.02(1)\\
$B(\mu_{h_3})$&1.48&1.51(1)&1.67&1.26(1)\\
\hhline{||-||-|-||-|-||}
$B(\mu_{h_1})$&1.41&1.04(1)&1.56&0.85(1)\\
$B(\mu_{h_2})$&1.72&1.26(1)&1.92&1.04(1)\\
$B(\mu_{h_3})$&2.11&1.52(1)&2.37&1.28(1)\\
\hhline{||-||-|-||-|-||}
$B(\mu_{h_1})$&2.02&1.08(1)&2.23&0.89(1)\\
$B(\mu_{h_2})$&2.46&1.30(1)&2.74&1.08(1)\\
$B(\mu_{h_3})$&3.01&1.55(1)&3.39&1.31(1)\\
\hhline{||-||-|-||-|-||}
$B(\mu_{h_1})$&2.50&1.12(1)&2.76&0.92(2)\\
$B(\mu_{h_2})$&3.05&1.34(1)&3.40&1.11(2)\\
$B(\mu_{h_3})$&3.73&1.58(1)&4.21&1.34(1)\\
\hhline{||-||-|-||-|-||}
$B(\mu_{h_1})$&2.92&1.16(1)&3.23&0.95(2)\\
$B(\mu_{h_2})$&3.56&1.38(1)&3.97&1.15(2)\\
$B(\mu_{h_3})$&4.36&1.60(2)&4.91&1.38(2)\\
\hhline{||-||-|-||-|-||}
$B(\mu_{h_1})$&3.66&1.25(1)&4.04&1.00(3)\\
$B(\mu_{h_2})$&4.46&1.46(1)&4.97&1.22(3)\\
$B(\mu_{h_3})$&5.46&1.66(2)&6.15&1.45(1)\\
\hhline{|b:=:b:==:b:==:b|}
\end{tabular}
\end{center}
\caption{\label{tabmassenergies}\em\footnotesize Masses and energies extracted from 
the two-point correlation functions in units of the lattice spacing. At $\beta=3.9$,
 time intervals for the fits are [8, 23] ($D$),
[6, 9] ($D^*_0$ and $D^*_2$),
[11, 17] (small momenta, $B(\mu_{h_1})$ and $B(\mu_{h_2})$), [9, 15] (large momenta, 
$B(\mu_{h_1})$ and $B(\mu_{h_2})$) and [9, 13] ($B(\mu_{h_3})$).
At beta=4.05, time ranges for the fits are [10, 26] ($D$), 
[7, 11] ($D^*_0$ and $D^*_2$),
[14, 26] (small momenta, $B(\mu_{h_1})$ and $B(\mu_{h_2})$), [9, 26] (large momenta, 
$B(\mu_{h_1})$ and 
$B(\mu_{h_2})$), [14, 22] (small momenta, $B(\mu_{h_3})$) and [9, 22] 
(large momenta, $B(\mu_{h_3})$). $\beta=4.2$ has been added to study the zero recoil
decay $B \to D^*_0$. The windows are [11,20] ($D$), [11,17]  ($D^*_0$) and [13,20] 
($B$ mesons at rest).}
\end{table}


\end{document}